# Magnetic ordering of the martensite phase in Ni-Co-Mn-Sn-based ferromagnetic shape memory alloys

Sudip Kumar Sarkar[1], Sarita Ahlawat[1], S. D. Kaushik[2], P. D. Babu[2], Debasis Sen[3], Dirk Honecker[4] and Aniruddha Biswas[1]

[1]Glass and Advanced Materials Division, Bhabha Atomic Research Centre, Mumbai - 400085, India

[2]UGC-DAE Consortium for Scientific Research, Mumbai Centre, BARC Campus, Mumbai - 400085, India

[3]Solid State Physics Division, Bhabha Atomic Research Centre, Mumbai - 400085, India

[4]Institut Laue-Langevin, Grenoble - 38042, France

E-mail: s.sudip.iitg@gmail.com

**Abstract:**

The magnetic state of low temperature martensite phase in Co-substituted Ni-Mn-Sn-based ferromagnetic shape memory alloys (FSMAs) has been investigated, in view of numerous conflicting reports of occurrences of spin glass (SG), superparamagnetism (SPM) or long range anti-ferromagnetic (AF) ordering. Combination of DC magnetization, AC susceptibility and small angle neutron scattering (SANS) studies provide a clear evidence for AF order in martensitic phase of $Ni_{45}Co_5Mn_{38}Sn_{12}$ alloy and rule out SPM and SG orders. Identical studies on another alloy of close composition of $Ni_{44}Co_6Mn_{40}Sn_{10}$ point to presence of SG order in martensitic phase and absence of SPM behavior, contrary to earlier report. SANS results do show presence of nanometre-sized clusters but they are found to grow in size from 3 nm at 30 K to 11 nm at 300 K, and do not correlate with magnetism in these alloys.

## 1. INTRODUCTION

Ni-Mn-Sn-based Heusler alloys and its Co-doped variants belong to a set of new Ga-free FSMAs that are known for their inverse magnetocaloric effect and have attracted attention of a large number of researchers in recent times due to fundamental interest and their strong application potential in magnetic refrigerator, sensors, actuators and energy conversion devices [1-5]. The strong magneto-structural coupling near martensitic transformation (MT) results in multifunctional properties like giant magnetocaloric effect (GMCE) and giant magneto resistance (GMR) in these materials [6-8]. In addition, these alloys show several novel magneto-structural phenomena such as, magnetic field induced reverse transformation (MFIRT), sometimes termed as meta-magnetic transformation [9-10] and kinetic arrest (KA) effect [11-12]. Magnetic field induced large change in magnetization from austenite phase with higher saturation magnetization to martensite with lower magnetization during transformation determines the extent of multifunctional effects in these alloys [13-14].

Substitution of Ni by Co began with the effort to minimize hysteresis in these alloys. Starting from full Heusler $Ni_2MnSn$ alloy, Co-substitution for Ni site in $Ni_{50-x}Co_xMn_{25+y}Sn_{25-y}$ alloys by Srivastava *et al* [15] resulted in an austenite phase with very large saturation magnetization (1170 emu/cm$^3$) with low magneto-crystalline anisotropy and a non-ferromagnetic martensite phase with very low thermal hysteresis (6 K) for the alloy with composition of $Ni_{45}Co_5Mn_{40}Sn_{10}$. In addition, Co substitution enhanced MT temperature to 410 K, which was well above room temperature. This combination of properties is particularly promising for applications. Following this, a seminal work on Co-doped Ni-Mn-Sn alloys was carried out by Cong *et al* [13], where a magnetic phase diagram was constructed in $Ni_{50-x}Co_xMn_{25+y}Sn_{25-y}$. It was shown that magnetism changed rapidly from a combination of paramagnetic austenite and anti-ferromagnetic (AF) martensite (x < 5) to a situation where austenite was ferromagnetic (FM) and martensite was non-ferromagnetic (AF/Paramagnetic) (x > 5). Magnetism becomes particularly complex in the region of 5 – 8 at. % Co. Absence of large spontaneous magnetization, irreversibility in Zero Field Cooling (ZFC)-Field Cooled Cooling (FCC) *M* versus *T* data, occurrence of exchange bias, frequency dependency of AC susceptibility peak, suggest that FM and AF exchange interactions are in strong competition. This leads to many possible magnetic ground state structures that involve spin clusters, like superparamagnetism (SPM), reentrant spinglass (RSG), conventional superspinglass/canonical spinglass (SSG), cluster spinglass (CSG), or alternately long range AF ordering. Similar low temperature behaviour has been noticed in a few other FSMAs as well. Long range AF ordering has been noticed by Aksoy *et al* for ternary $Ni_{50}Mn_{37}Sn_{13}$ and $Ni_{50}Mn_{40}Sb_{10}$ system [16-17]. RSG behavior is established for bulk $Ni_{1.6}Mn_2Sn_{0.4}$ alloy by Ma *et al* [18] and for $Ni_2Mn_{1.36}Sn_{0.64}$ alloy by Chatterjee *et al* [19] while $Ni_{1.6}Mn_2Sn_{0.4}$ ribbon shows CSG behaviour as reported by Singh *et al* [20]. CSG ground state has also been reported by Tian *et al* for $Ni_{50}Mn_{38}Ga_{12}$ and $Ni_{50}Mn_{38}Ga_{11}Sb_1$ alloys [21]. SSG behaviour are noted by Wang *et al* in $Ni_{50}Mn_{37}In_{13}$ [22], Umetsu *et al* in $Ni_{50}Mn_{40}Sb_{10}$ [23], Srivastava *et al* in $Ni_{55}Mn_{21}Al_{22}$ [24], and by Liao *et al* in $Ni_{50}Mn_{36}Co_4Sn_{10}$ alloys [25]. CSG has been probed for $Ni_{50}Mn_{34.8}In_{15.2}$, $Ni_{50}Mn_{38.5}Sn_{11.5}$ alloys by Umetsu *et al* [23], for $Ni_{50}Mn_{38}Ga_{12}Sb_x$ (x= 3, 4, 5, 6) by Tian *et al* [21] and for $Ni_{50}Mn_{34}Sn_6Al_{10}$ alloy by Agarwal *et al* [26]. A transition from SPM state to SSG has been reported by Cong *et al* in a Co-doped quaternary Ni-Mn-Sn alloy [13]. In another similar work, Cong *et al* has shown SSG ground state below spin freezing temperature and SPM state above freezing temperature for $Ni_{43.5}Co_{6.5}Mn_{39}Sn_{11}$ alloy [27]. SPM spin freezing is observed by Perez-Landzabal *et al* in $Ni_{45}Co_5Mn_{36.7}In_{13.3}$ [28] and by Bhatti *et al* in $Ni_{44}Co_6Mn_{40}Sn_{10}$ and $Ni_{52}Co_8Mn_{40}Sn_{10}$ alloys [29]. With the exception of [29], all the above studies utilized AC susceptibility analysis to characterize the magnetic structures that involved spin clusters. SANS is another particularly effective technique in studying steady state structure of spin cluster owing to its ability to probe the mesoscopic density fluctuations. Additionally, SANS is capable to provide quantitative dimensional analysis of the spin clusters. Bhatti *et al* [29] is the only one who have employed SANS to investigate FSMAs and their analysis revealed a liquid like spatial distribution of interacting magnetic clusters of 12 nm for $Ni_{44}Co_6Mn_{40}Sn_{10}$ and 14 nm for $Ni_{42}Co_8Mn_{40}Sn_{10}$ alloys. They also reported SPM freezing of spin clusters in $Ni_{44}Co_6Mn_{40}Sn_{10}$ alloy. One particularly striking observation in their study is the growing size of the spin clusters with increasing temperature. This contradicts the widely accepted view that spin clusters shrink with increasing temperature, as thermal energy weakens inter spin interactions [30-32]. In fact, this brings us to the following two important points: (i) it is important to eliminate the possibility of occurrence of nanometric structural clusters during such SANS analysis; (ii) it is prudent to combine SANS analysis with AC susceptibility measurement in order to conclusively identify the nature of the clusters.

In view of this, the current study focuses on a deeper understanding of magnetic correlations in low temperature martensite phase of Co-substituted Ni-Mn-Sn system, mainly $Ni_{45}Co_5Mn_{38}Sn_{12}$ and for the sake of parity $Ni_{44}Co_6Mn_{40}Sn_{10}$, using a combination of complimentary techniques like SANS and temperature dependent detailed magnetometry including AC susceptibility. We have used neutron diffraction and transmission electron microscopy for structural analysis, and differential scanning calorimetry to determine the transformation temperatures. Interestingly and contrary to earlier reports, we find no experimental evidence for spin clusters in our study. Even though SANS analysis clearly detects nanometric-clusters in $Ni_{45}Co_5Mn_{38}Sn_{12}$ alloy, they turn out to be structural in nature. Concurrent AC susceptibility experiment helps ascertain this. Thus, no evidence for any other effect like SG or SPM is present in this alloy. Temperature dependent AC susceptibility reveals long range AF ordering in the low-temperature martensite of $Ni_{45}Co_5Mn_{38}Sn_{12}$, whereas SSG state is found in case of $Ni_{44}Co_6Mn_{40}Sn_{10}$, which is also at variance with the earlier report.

## 2. EXPERIMENTAL

The buttons with nominal compositions of $Ni_{45}Co_5Mn_{38}Sn_{12}$ and $Ni_{44}Co_6Mn_{40}Sn_{10}$ were prepared by vacuum arc melting high purity (99.99 %) elements in appropriate proportions. Homogeneity was ascertained by re-melting the alloys multiple times. The buttons were sealed in a quartz ampoule filled with helium gas, solutionized at 1123 K for 24 h, and quenched in ice water. Detailed characterization of these alloys was carried out using scanning & transmission electron microscopy (SEM & TEM), energy dispersive x-ray spectroscopy (EDS), x-ray diffraction (XRD), neutron diffraction (ND), differential scanning calorimetry (DSC), DC magnetization, AC susceptibility and SANS techniques. Samples for metallography were etched using an aqueous solution of $FeCl_3$ in HCl. XRD experiments were carried out for both bulk and powder samples using a Cu $K_\alpha$ radiation. ND experiments were carried out using a focusing crystal diffractometer at Dhruva reactor, BARC, India at 1.48 Å wavelength of neutron beam. Specimens for TEM were prepared by slicing discs from an electro-discharge machined cylindrical rod of 3 mm diameter, followed by grinding and jet polishing with a Struers Tenupol-5 at 233 K, using a 10 vol.% perchloric acid in methanol electrolyte. TEM was performed employing a JEOL 2000FX microscope at 200 keV. Isochronous DSC experiments were performed using a Mettler-Toledo calorimeter at a rate of 10 K/min in argon atmosphere. Magnetometry was carried out using a commercial 9 Tesla PPMS-VSM (by Quantum Design). SANS experiments were performed at the D11 SANS facility at the Institute Laue-Langevin (ILL), Grenoble, France. For the in-situ SANS measurement, an as-solutionized sample was installed on an Orange cryostat. In-situ experiment was performed at several key transformation temperatures starting from 30 K to 300 K. The incident beam was collimated by a rectangular aperture of 10 mm x 7 mm and another Cd aperture, 5.5 mm in diameter, was mounted in front of each sample to define the illuminated area. An incident beam with wavelength $\lambda$ = 6 Å (spread in wavelength: 10% of full width at half maximum) was used. The accessible wave-vector transfer $q$ covers the range 0.002 to 0.125$Å^{-1}$, after combining data from two different sample-to-detector distances (1.3 and 8.0 m). Raw data were corrected for background, transmission and electronic noise, and converted to absolute macroscopic differential scattering cross sections.

## 3. RESULTS

### *3.1. $Ni_{45}Co_5Mn_{38}Sn_{12}$ alloy:*

#### *3.1.1. Microstructure:*

A two-phase microstructure is observed in as-solutionized condition for $Ni_{45}Co_5Mn_{38}Sn_{12}$ alloy as shown in a representative SEM image in figure 1. Typical grain size is around 400 µm. The microstructure at room temperature comprises mainly austenite matrix and some fraction of martensite. This is expected because martensite start temperature is below but close to room temperature. The inset of figure 1 shows the representative SEM image for the zoomed portion of the twinned martensite region. The chemical composition of this alloy, determined by EDS attached to an SEM, is listed in table 1. Experimentally determined composition is very close to the nominal alloy composition. The errors presented in the composition analysis correspond to one standard deviation. Valence electron concentration per atom ratio (*e/a*) is also computed for this alloy and is included in table 1.

#### *3.1.2. Thermal analysis:*

Figure 2 shows the isochronal DSC plots of $Ni_{45}Co_5Mn_{38}Sn_{12}$ in as-solutionized condition. Clear evidence of reversible structural martensitic transformation is observed. Curie transition, as evident in the DSC scans and is characterized by a change in the base line. Both the austenite Curie temperature ($T_{CA}$) and martensite Curie temperature (inset of figure 2) ($T_{CM}$) are detected for $Ni_{45}Co_5Mn_{38}Sn_{12}$ alloy.

Characteristic MT temperatures such as, martensite start temperature ($M_s$), martensite finish temperature ($M_f$), austenite start temperature ($A_s$) and austenite finish temperature ($A_f$) along with $T_{CA}$ and $T_{CM}$ are listed in table 2 for this alloy.

### *3.1.3. ND and TEM analysis:*

Figures 3(a)-(b) display ND patterns of $Ni_{45}Co_5Mn_{38}Sn_{12}$ alloy at 300 K and 4 K, respectively. Figure 3(a) shows the Rietveld fitted ND pattern at room temperature (300 K) of the austenitic phase having $L2_1$ structure of *Fm*-3*m* space group. The result is consistent with the XRD study carried out by Jing *et al* on the same composition [33]. Figure 3(b) corresponds to the martensite phase. The full profile pattern was fitted using Le Bail method for martensite phase. Martensite is 6M monoclinic type having $P2_1$ space group. Figure 4(a) shows the bright-field TEM micrograph of $Ni_{45}Co_5Mn_{38}Sn_{12}$ alloy that depicts fine martensite plates. Corresponding Selected Area Electron Diffraction (SAED) pattern is in figure 4(b) that clearly shows evidence for 6-layer modulation in terms of the satellite spots.

### *3.1.4. Magnetization measurement:*

Figure 5(a) shows magnetization as a function of temperature (*M* versus *T* plot) in the range 5-400 K for applied magnetic field of 100 Oe for $Ni_{45}Co_5Mn_{38}Sn_{12}$ alloy in ZFC, FCC and field-cooled warming (FCW) sequences. On cooling from 400 K, there is an abrupt increase in magnetization at around 375 K, which corresponds to $T_{CA}$, where paramagnetic austenite transforms to ferromagnetic austenite. On further cooling, magnetization drops suddenly at 292 K, which corresponds to the martensitic transformation of the alloy. On the warming cycle, sudden raise of magnetization seen at 278 K corresponds to reverse martensitic (martensite to austenite) transformation. A thermal hysteresis observed between FCC and FCW at around 276

K to 292 K is the characteristics of first order martensitic transformation. Magnetization starts to increase below 225 K, and the field cooled (FCC & FCW) curves tends to saturate at certain value at low temperatures. On the other hand, the ZFC curve goes through a maximum, termed as spin freezing temperature ($T_f$), at around 165 K, and then drops nearly to zero (very small value) as $T$ reaches 5 K. A significant bifurcation between ZFC and FCC/FCW curves is observed below 165 K. Similar behavior was seen earlier in other related alloys such as $Ni_{50}Mn_{25+y}Sn_{25-y}$ [34], $Ni_{50}Mn_{25+y}In_{25-y}$ [35-36], and $Ni_{50-x}Co_xMn_{40}Sn_{10}$ [27, 29, 37], and such a behavior has been interpreted as SPM or SG or AF ordering in the existing literature.

Figure 5(b) displays representative $M$ versus $H$ plots for $Ni_{45}Co_5Mn_{38}Sn_{12}$ alloy at selected temperatures. Above $T_{CA}$, for example at 399 K, the variation of $M$ versus $H$ curve is typical of a paramagnet, thereby clearly indicating austenite paramagnetic region. The $M$ versus $H$ curves at 370 K and 340 K, i.e., below $T_{CA}$, exhibit typical FM behavior. The austenite FM region extends from temperature span of $T_{CA}$ down to $M_s$. In fact, there is no confusion on the magnetic nature of austenite in Ni-Mn-Sn based Heusler alloys. Similarly, $M$ versus $H$ curves in the region of martensitic transformation, i.e., around 278–305 K, exhibit field-induced reverse martensitic phase transformation, which is well known. The $M$ versus $T$ curves of figure 5(a) seem to suggest that 225 K is the $T_{CM}$ and it matches with the value as obtained from DSC. However, the $M$ versus $H$ curve (figure 5(c)) at 225 K indicates a weak magnetic order with a spontaneous magnetization of ~9emu/g and a coercive field of 200 Oe. The origin of this magnetic order at 225 K will be discussed later.

Figure 5(c) shows the full $M$ versus $H$ plot for temperature above $T_f$ as shown for three selected key temperatures of 225, 230 and 260 K. None of the curve follows sigmoidal behaviour which is the characteristics of SPM ground state. In fact, non-linear variation with field with presence of hysteresis for the curves clearly rules out blocked SPM ground state above $T_f$. Inset of Figure 5(c) shows the variation of coercivity with temperature for the alloy. Coercive field ($H_C$) can be estimated from the $M$ versus $H$ loops as $(H_{C1} + H_{C2})/2$ where $H_{C1}$ and $H_{C2}$ are coercive fields in the positive and negative field axis [38-39]. $H_C$ initially increases with temperature and begins to decrease after reaching a maximum value. This typical variation of $H_C$ with temperature can be explained from the anisotropy of AF ordering present in the alloy. It is known that anisotropy of AF region decreases with increasing temperature, because AF region is able to drag more spins resulting in increasing $H_C$ below $T_f$ [21].

Figure 5(d) depicts the magnified portion of the low field (0 to 3 kOe) *M* versus *H* curves over a wide temperature range of 5 K to 225 K. It clearly shows AF to FM meta-magnetic transition in all the virgin curves for $T < 180$ K. This meta-magnetic transition indicates that the ground state is AF and field drives the system to FM. This transition is more pronounced at low temperatures and the field at which it occurs ($H_{cr}$), decreases progressively with increasing temperature. It is barely noticeable at 160 K and 180 K. Similar AF ordering of martensitic phase is also seen earlier in Ni-Mn-based FSMAs [16-17]. However, this observation is in contradiction with the report of Bhatti *et al* [29] on $Ni_{45}Co_5Mn_{40}Sn_{10}$ alloy, where they have interpreted such behavior as evidence for SPM state.

We, therefore, carried out additional analysis to test for the presence of SPM as well, despite evidence pointing to AF order. One of the first signatures for SPM is that generally the FC curve should show an increasing trend as *T* approaches base temperature instead of showing a saturating trend like in the present case [21]. A better test for SPM behavior is that the magnetization isotherms should collapse on to a single universal curve in a plot of *M* versus *H/T*, above the blocking temperature [40-41]. Figure 6 shows such a plot of *M* versus *H/T* for $Ni_{45}Co_5Mn_{38}Sn_{12}$ alloy in the temperature range spanning the blocking temperature. It is clearly seen that none of the curves collapse on to a single universal curve. In addition to that, the Arrott's plot (inset of figure 6) demonstrates the presence of spontaneous magnetization as evident from the positive intercept of the extrapolated high field $M^2$ versus *H/M* curves [42-45]. These observations clearly rule out SPM order in $Ni_{45}Co_5Mn_{38}Sn_{12}$ alloy.

The frequency dependent AC susceptibility results provide further valuable information about the SG and/or SPM behaviour in magnetic materials. We performed AC susceptibility measurements on $Ni_{45}Co_5Mn_{38}Sn_{12}$ at an ac driving field of 7 Oe and at several selected frequencies up to a maximum of 10 kHz over a wide range of temperature. Figure 7(a) shows the temperature dependence of the real part of AC susceptibility ($\chi'$) at $f$ = 93, 193, 993, 1111, 5533 and 9984 Hz, and the inset shows the magnified portion around 175 K. The observed peak in $\chi'$ plot corresponds to the one that is seen in the ZFC magnetization curve (figure 5(a)). We note that the peak in ZFC is at ~ 165 K compared to the one at ~175 K in $\chi'$. This difference is not unusual if we consider the following: (i) ZFC is measured in 100 Oe DC field ($H_{DC}$) whereas $\chi'$ in 7 Oe ac field and $H_{DC} = 0$; and (ii) with increasing DC fields the ZFC peak progressively shifts to lower temperatures. The peak at $T_f$ in $\chi'$ versus T curve would be quite sensitive to the frequency of applied ac driving field. Both for SG and SPM systems, this peak at $T_f$ is expected to shift towards higher temperatures and the peak height is expected to decline with increasing frequency. The extent of peak shift with frequency is different for SG and SPM systems and can be quantitatively analyzed in terms of the following expression [46-47]:

$$\Phi = \Delta T_f / (T_f \, \Delta log_{10} f) \quad (1)$$

where, $\Phi$, sometimes termed as Mydosh parameter, represents the relative shift of $T_f$ per decade of frequency change. The value of $\Phi$ decreases with increase in inter-particle interaction strength. Such an analysis results value of $\Phi$= 0.0017 for the $Ni_{45}Co_5Mn_{38}Sn_{12}$ alloys which is neither in agreement with that of the SG system ( $\Phi$ ~ 0.005 to 0.01) [31,47] nor with non-interacting ideal SPM systems ( $\Phi$~ 0.1) [32, 48]. SG systems are further classified into CSG and SSG based on $\Phi$ value as well. For a CSG system the value of $\Phi$ is 0.01 [49] while for conventional SSG it is 0.001 [50]. For a well-ordered FM or AF system the value is almost zero [49]. Therefore, it is evident that the alloy $Ni_{45}Co_5Mn_{38}Sn_{12}$ lies in the borderline of SSG and long range AF/FM ordered ground state, hence, required further test.
The frequency dependence of the glassy systems should also follow conventional power-law divergence by critical slowing down (CSD) model [48, 51]:

$$\tau = \tau_0 (T_P/T_f - 1)^{-zv} \quad (2)$$

where $\tau$ is relaxation time corresponding to the measured frequency ($\tau = 1/f$), $\tau_0$ is single spin flip relaxation time, $T_f$ is the freezing temperature corresponding to zero frequency ($f$ = 0) and $zv$ is the dynamic critical exponent. The parameters were obtained from the best fit of experimental data using equation (2), depicted as ln($f$) versus $1/(T_f$-$T_0)$ plot as shown in figure 7(b). The estimated value of $zv$ turns out to be 249.7 and $\tau_0 \approx 8.34\times10^{-43}$ s. The value of $zv$ is unrealistically large (for spin glass system $2 < zv < 10$) whereas $\tau_0$ is abnormally small (for spin glass system, $10^{-7} < \tau_0 < 10^{-14}$ s) [31-32].

Another dynamical law to characterize weakly interacting glassy systems is the Vogel-Fulcher (V-F) law [48, 52-53]:

$$\omega = \omega_0 exp\left(-E_a/K_B(T_f - T_0)\right) \quad (3)$$

where $\omega$ (= $2\pi f$) is the measurement angular frequency, $\omega_0$ is the characteristic frequency of the spin glass, $E_a$ is the activation energy of the spin glass, $K_B$ is the Boltzmann constant, and $T_0$ is the V-F temperature that describes the interaction among spin clusters. The values of $T_0$ are very close to the freezing temperature corresponding to zero frequency as obtained from DC magnetization and $\tau_0$ (= $2\pi/\omega_0$) is the microscopic relaxation time related to spin glass or cluster spin glass. The parameters were obtained from the linear fitting of ln($f$) versus 1/($T_f$-$T_0$) as shown in the upper inset in figure 7(b). The best fit yields, $E_a/K_B$ = 1.30×10$^3$ K and $\tau_0$ = 8.24×10$^{-46}$ s. These parameters are unphysical and not in agreement with the values reported in the literature for spin glass. For example, $\tau_0$ varies from 10$^{-11}$ - 10$^{-15}$ s for SSG, and the typical value for CSG compounds: $\tau_0$= 10$^{-9}$ s [19, 26, 31]. Therefore, all the empirical parameters obtained from fitting of AC susceptibility data definitely rule out the possibility of spin glass ordering in $Ni_{45}Co_5Mn_{38}Sn_{12}$ alloy.

We have also investigated for the possibility of the presence of non-interacting spin cluster in case of SPM ground states. For non-interacting isolated spin clusters, the relaxation time ($\tau$) follows the Neel- Arrhenius (N-A) model as follows [54-55]:

$$\tau = \tau_0 exp\left(E_a/K_B T_f\right) \qquad (4)$$

where $\tau_0$ is the attempt time with typical value of 10$^{-9}$-10$^{-10}$ s [51] and $E_a$ is the anisotropy energy barrier. An attempt was made to fit the experimental data using equation (4) and the best fit yields $\tau_0$=1.58×10$^{-99}$ s which is physically an unrealistic quantity and thus rules out the possibility of SPM state. The experimental data along with the best fit are depicted as ln($f$) versus 1/$T_f$ plot, shown in lower inset of figure 7(b). The detailed extracted parameters from Mydosh, CSD, V-F and N-A analyses are listed in Table III.

### *3.1.5. Small-Angle Neutron Scattering:*

In this experiment, total scattering cross-section was measured as a function of scattering wave vector, $q$, at five specific temperatures (30 K to 300 K), representative of the various important temperature regimes of phase transformation. The data were taken in zero applied magnetic field, starting from 30 K, in the heating mode. The scattering was isotropic in the scattering plane and the 2D scattering data were radially averaged to obtain 1D scattering cross section (Intensity, $I$) versus $q$.

A close inspection of the scattering data at 30 K, as shown in figure 8(a), reveals that there are essentially three contributions to the scattering. At low $q$ (below about 0.006 Å$^{-1}$), the data are well described by a straight line in double logarithmic representation, indicating power-law of scattering. While in the high $q$ region of above about 0.006 Å$^{-1}$, the scattering data could not be described by simple Porod's power law scattering, but can be fitted with Lorentzian function. In addition, and most interestingly, a peak appears at $q$ value around 0.05 Å$^{-1}$, which contains wealth of information about the formation of nano-clusters. The peak can be fitted as a Gaussian function. The overall scattering intensity is described as [29]:

$$\frac{d\Sigma(q,T)}{d\Omega} = \frac{\left(\frac{d\Sigma}{d\Omega}\right)_P(T)}{(q^n)} + \left(\frac{d\Sigma}{d\Omega}\right)_G(T)e^{-[(q-q_G(T))^2/(2\Delta(T)^2)]} + \frac{\left(\frac{d\Sigma}{d\Omega}\right)_L(T)}{(q^2+(1/\xi)^2)} \quad (5)$$

where the first term on the right hand side of equation (5) is Porod's power law contribution while second and third term corresponds to Gaussian and Lorentzian contributions, respectively, to the total intensity. Individual Porod, Gaussian and Lorentzian contributions along with overall (sum of three contributions) fits are shown in the figures 8(a) - (c) for three selected scattering data (30 K, 243 K, 300 K) as an example. The physically relevant parameters are extracted from the fits and their variation with $T$ is shown in figure 8(d).

Now, coming to Porod's power law of scattering, *i.e.*, first in equation (5), $(d\Sigma/d\Omega)_P$ is a constant that gives the strength of the Porod's contribution and $n$ is an exponent that provides information about the nature of the scattering centres. The first term in equation (5) is valid in the limit $q \ll 2\pi/d$, where $d$ is the size of the scattering object. For 30 K data in our case, $n$ was found to be 4.18, which suggests scattering from an assembly of three-dimensional (3D) objects with "smooth" surfaces. As temperature is increased, value of $n$ increases slightly until the martensitic transition region is reached, where $n$ value shoots up to 5.6. As encountered previously in a wide variety of materials, we interpret this as scattering from grains and grain boundaries. The adherence to this form down to $q$ = 0.002Å$^{-1}$ indicates that qualitatively these grains are larger than 100 nm in size. These nanometric grains can be magnetic domains. The domain size increases with increase in temperature and it becomes diffuse while reaching 300 K when the alloy is above the $A_s$ temperature and below the $A_f$ temperature. The second term in equation (5) is the Gaussian term, described by $q_G$ (position of peak) and $\Delta$, width of peak, associated with size and distribution of nano-cluster while $(d\Sigma/d\Omega)_G$ is the strength of Gaussian scattering. The $\Delta$ in fact signifies spatial correlation of the clusters. The values of $q_G$ decreases with increase in temperature. Similarly, $\Delta(T)$ shows slight decrease in value from 30 K to 150 K and very slow increase till 300 K, demonstrating that spatially clusters correlation does not change with temperature. In other words, the result rules out SG where cooperating spin freezing is essential [56]. The result is in line with detailed magnetization study. The third term in equation (5) is the Lorentzian term. Though the Lorentzian term has slight contribution in intensity, the spin correlation length $\xi$ is too low in the temperature region 30-300 K and it is consistent with earlier report that $\xi$ is significant only above $T_{CA}$ temperature (375 K) [29]. Inset of figure 8(d) shows the variation of scattering intensity with temperature for $q$ = 0.002 Å$^{-1}$. The scattering intensity decreases first with increase in temperature up to 190 K after which it increases with temperature till 300 K. At this $q$ value, scattering signal comes from magnetic domain. The observation is consistent with low temperature magnetic ordering of the alloy as revealed from magnetization study. The alloy remains AF till 225 K but its ordering weakens with increase in temperature: so scattering intensity first decreases and above 225 K, the alloy attains FM order that shows increasing exchange interaction with temperature. As a result, scattering intensity shows increase above 190 K due to presence of long range FM domains. Scattering results are consistent with magnetism results.

It is now time to look into the most significant part of the scattering data, i.e., the appearance of peak at higher $q$ value: for example, at ~ 0.05Å$^{-1}$ for the 30 K data. This peak shifts towards lower $q$ with increase in temperature and becomes progressively more prominent. The scattering profiles have been fitted using a model of assemblies of spherical scattering objects. The scattering intensity can be described by the following equation under local monodisperse approximation (Pedersen *et al*) [57]:

$$I(q) = n_t \Delta\rho^2 \int P(q,r) N(r) V_p^2(r) S(q,r) dr \qquad (6)$$

Here, $n_t$ is the number density of the nano-clusters, $\Delta\rho^2$ is the contrast factor, determined by the difference between the scattering length densities of nano-cluster and matrix. In equation (6), $P(q,r)$ is the single particle form factor, which is a function of the cluster shape and size.

For spherical cluster of radius $r$, the form factor is given as [58]:

$$P(q,r) = 9\left[\frac{\sin(qr) - qr\cos(qr)}{(qr)^3}\right]^2 \qquad (7)$$

The cluster size distribution is fitted to a log normal distribution [59]:

$$N(r) = \frac{N}{r\sigma\sqrt{2\pi}} e^{-\frac{[\ln(r/\mu)]^2}{2\sigma^2}} \qquad (8)$$

where, r is the median cluster radius, μ is the standard deviation, N is a normalization factor and σ is the polydispersity.

In the case where the particles are dilute, or for magnetic scattering, if the magnetic anisotropy is randomly oriented [60], peaks in *I(q)* versus *q* plot arise from *P(q)* and the structure factor can be assumed as unity. No form factor oscillations are observed presumably due to the large polydispersity present in the system. The peak as well as whole scattering data were fitted keeping $n_t$ and $\Delta\rho$ (nuclear) constant using equation (6) to estimate the size of the nano-clusters. The complete scattering data was fitted by considering two kinds of spherical particles: the large one corresponds to magnetic domain that gives rise scattering at low *q* values and a population of nano-cluster with an associated correlation peak. Using this model, 30 K and 150 K scattering data was fitted while another intermediate particle size needs to be introduced for the satisfying fits of 190 K, 243 K, 280 K and 300 K scattering data. Figure 9(a) shows the experimental scattering data with the fits using equation (6). The whole scattering data are divided into three regions. Region-I corresponds to scattering from magnetic domains (50-113 nm) while region-II corresponds to scatterers of 11-13 nm size and region-III for very small clusters (3-4 nm). The scattering objects from regions-II and III are correlated to martensitic transformation because scattering becomes stronger as it approaches martensitic transformation region. The small peak at 30 K develops to a broad peak at 280 K (which is the $A_s$ temperature) and is most prominent feature at 300 K. This result clearly indicates that the origin of these nanometric structures is related to structural martensitic transformation and these nano-clusters could be untransformed austenite. The size information extracted from the fits for particles associated with scattering from regions-I, II and III are shown in figure 9(c).

To have the information of the volume fraction ($\varphi$) of the associated structural nanometre-sized clusters, integrated intensities were calculated from the $I(q)q^2$ versus *q* plot as shown in figure 9(b) below. The integrated intensity can be calculated as the area under the $I(q)q^2$ versus *q* plot as [61-62]:

$$Q = \int_0^\infty I(q)\, q^2 dq \qquad (9)$$

The value of *Q* depends on the volume fraction and the scattering contrast of the nano-cluster with matrix according to the following equation:

$$Q = 2\,\pi^2\, (\Delta\rho)^2\, \varphi\, (1-\varphi) \qquad (10)$$

Since $\varphi$ is a very small term, $\varphi^2$ is ignored, as compared to $\varphi$. So, the integrated intensity is proportional to the volume fraction ($\varphi$) of nano-clusters as the scattering contrast ($\Delta\rho$) is constant. Separate volume fraction is calculated for clusters for regions-II and III. Inset of figure 9(c) shows the plots of the relative variation of volume fraction with temperature for region-II and III. It is obvious that the volume fraction related to the structural cluster increases in region-II while it decreases with increase in temperature for region-III, which in turn suggests that the clusters grow in size at the expense of smaller clusters with rise in temperature.

### *3.2. $Ni_{44}Co_6Mn_{40}Sn_{10}$ alloy:*

Interestingly, the only other SANS study on FSMAs was by Bhatti *et al* [29] in Ni-Co-Mn-Sn alloys who reported SPM freezing of spin clusters for the compositions: $Ni_{44}Co_6Mn_{40}Sn_{10}$ and $Ni_{42}Co_8Mn_{40}Sn_{10}$. Notwithstanding the difference in interpretation, it is worthwhile to note that there is striking similarities between their SANS results, and what we observe in our experiments. Their scattering data also show a peak interpreted to nanometric clusters and those clusters too grow in size with increasing temperature. However, detailed complimentary magnetometry helps us conclusively decipher the low temperature magnetic order in our alloy $Ni_{45}Co_5Mn_{38}Sn_{12}$, which is having AF ground state, if at all. This obvious contradiction has driven us to re-examine the alloy of their focus namely, $Ni_{44}Co_6Mn_{40}Sn_{10}$, in terms of detailed magnetization investigation.

The first challenge was to experimentally achieve the same composition as $Ni_{44}Co_6Mn_{40}Sn_{10}$. table 1 shows the composition measured by EDS attached to SEM. High vapour pressure of Mn is responsible for the slight difference in the composition. Figure 10 displays the representative DSC scans for the alloy and table 2 shows the characteristic MT temperatures for the alloy which are very close to the values reported by Bhatti *et al* [29]. This result signifies that the alloy we have prepared is very similar to Bhatti *et al* [29], because MT temperatures are known to be extremely sensitive to composition.

Figure 11(a) depicts the $M$ versus $T$ plot for $Ni_{44}Co_6Mn_{40}Sn_{10}$ alloy. The bifurcation between ZFC and FC curves is seen from 300 K downwards. The ZFC data in the present case show a broad peak in ZFC curve centered around 100 K, while Bhatti *et al* observed similar broad peak around 60 K in ZFC data at $H$ = 10 Oe. The $M$ versus $H$ data in the low field region shown in the figure 11(b) exhibit a weak S-shaped AF-FM transition like curves for temperatures 5, 20, and 40 K, respectively, thereby indicating a weak AF order in this temperature range. Above 40 K, a FM behavior is observed as shown in figure 11(b). The variation of coercivity with temperature is shown in the inset of figure 11(b), which is similar to that of $Ni_{45}Co_5Mn_{38}Sn_{12}$ alloy. Figure 12 demonstrates that the magnetization isotherms for $Ni_{44}Co_6Mn_{40}Sn_{10}$ alloy do not collapse on to a single curve between blocking temperature ($T_f$ = 100 K) and MT temperature in the same way as $Ni_{45}Co_5Mn_{38}Sn_{12}$ alloy. This result is a clear indication that $Ni_{44}Co_6Mn_{40}Sn_{10}$ does not possess SPM in the specified temperature range between blocking and MT temperatures. The Arrott's plot of $M^2$ versus.*H/M* shown in the inset of figure 12 clearly demonstrates the presence of spontaneous magnetization, and hence suggests presence of long range magnetic order in the material.

Figure 13(a) shows the temperature dependence of the real part of AC susceptibility ($\chi'$) for $Ni_{44}Co_6Mn_{40}Sn_{10}$, measured at applied frequencies of 93, 193, 993, 1111, 5533 and 9984 Hz, respectively. Inset of figure 13(a) shows the clear frequency dependence of susceptibility peak. The frequency dependent susceptibility data were analyzed by applying Mydosh, CSD, V-F and N-A formalisms, exactly in the same way as described for $Ni_{45}Co_5Mn_{38}Sn_{12}$ alloy. The experimental data along with best fit by applying equation (2) for CSD model are shown in the form of ln($f$) versus ln($T_f/T_P$-1) plot in figure 13(b). Upper inset of figure 13(b) shows ln($f$) versus 1/($T_f$-$T_P$) plot for experimental data along with best fit by using equation (3) for V-F model. Experimental data along with best fit by using equation (4) for N-A model in the form of ln($f$) versus 1/$T_f$ plot is shown in the lower inset of figure 13(b). The parameters extracted from such analyses are listed in table 3.

As can be seen from the table 3, the result for $Ni_{44}Co_6Mn_{40}Sn_{10}$ are in well agreement with conventional SSG parameters and thus refuting earlier claim of SPM behavior in this compound.

## 4. DISCUSSIONS

Quaternary $Ni_{45}Co_5Mn_{38}Sn_{12}$ alloy presents predominantly austenite matrix with some fraction of martensite at room temperature, as corroborated by SEM and validated by thermal analyses. Austenite has $L2_1$ structure and martensite shows a 6M monoclinic one as confirmed by ND and TEM analyses. The central aim of the current work was to identify the true nature of the low temperature magnetic order in $Ni_{45}Co_5Mn_{38}Sn_{12}$ alloy using a combination of complementary techniques like SANS and temperature dependent magnetometry. At very low temperature (5 K), the martensite shows long range AF order, proven unambiguously by magnetometry and AC susceptibility. The magnetization study for the alloy shows AF ordering up to 225 K, a weak FM behaviour up to the MT temperature (280 K) and a strong FM order above the MT temperature. Variation of scattering intensity with temperature for $q = 0.002$ Å$^{-1}$, as obtained by the SANS data is consistent with this description of magnetic ordering of the alloy. The bifurcation between ZFC and FC magnetization below MT temperature and appearance of a broad peak in ZFC plot at 165 K, raises suspicion of a possible SG structure or SPM with a spin freezing or blocking temperature of 165 K. In fact, there are many conflicting reports in the literature on Ni-Mn-based Heusler type FSMAs where the alloys show spin glass nature, SPM nature or simple AF ordering at low temperature. Under such circumstances, frequency dependent AC susceptibility data are effective in establishing the true nature of magnetic ordering. The magnitude of Mydosh parameter obtained from the fits of AC susceptibility data is 0.0017 which is beyond the range of values reported for typical SG ($0.005 < \Phi < 0.05$) and even smaller than those for non-interacting ideal SPM systems ($\Phi \sim 0.1$) [47]. V-F parameter is also quite small compared to the typical value reported for interacting SSG [47]. The N-A parameter is found to be unrealistically small compared to the typical value for non-interacting SPM system [47]. Table 4 collates various spin dynamics scaling factors in Ni-Mn-based FSMAs from literature [13, 18-23, 25-27, 63] for better comparison. In addition, the presence of a meta-magnetic transition in all the virgin magnetization isotherms below 180 K also confirms that $Ni_{45}Co_5Mn_{38}Sn_{12}$ alloy indeed possesses long range AF ordering at low temperature.

On the other hand, SANS results show a peak at higher $q$ value of around 0.05Å$^{-1}$ for the entire range of temperature: 30 - 300 K. With increasing temperature, this peak gradually shifts towards lower $q$ and becomes increasingly prominent. In general, such thermal response is interpreted in the literature as an indication of magnetic origin of the peak [29], and presence of such peaks is explained in terms of nanometric FM spin clusters within AF matrix [29]. As regards the fundamental origin of spin clusters, there are many conflicting reports in the literature. Several researchers [28, 64] consider formation of Co-rich or Mn-poor regions in Ni-Co-Mn-Sn-based alloys as the main reason. In addition, magnetic phase separation in single chemical phase [29], electronic phase separation [29] and even untransformed austenite fraction [64-65] are shown as the origin of spin clusters. However, it is critically important to ensure the true identity of these nanometric clusters in the first place. In case of spin clusters, (i) size should decrease with increasing temperature as temperature is expected to weaken inter-cluster interaction; and (ii) they should cease to exist beyond $T_f$ [30-32]. But, SANS analysis in our case reveals quite the opposite. Clusters, in our study, show definite growth with increasing temperature. Moreover, they persist way beyond $T_f$, even beyond $A_s$ as well. If the clusters are of magnetic origin, they should have shown some significant changes across any of the magnetic transition. On the other hand, they continue to grow well into austenite phase and even beyond $T_{CA}$. This is true even for $Ni_{44}Co_6Mn_{40}Sn_{10}$ alloy reported by Bhatti *et al* [29]. Clearly it is difficult to attribute their origin to magnetic nature from the evidence that is presented so far. In all likelihood, they are of structural nature. One can reasonably assume these nanometric clusters to be untransformed austenite, which is expected to grow in size with increasing temperature. SANS data shows that there are, in fact, clusters having two different populations with 3-4 nm and 11-13 nm. Increase in temperature causes disappearance of small clusters and growth of bigger clusters. The relative volume fraction calculated from the integrated SANS intensity decreases with increasing temperature in region-III and increases in region-II as shown in inset of figure 9(c). So, it is evident that large cluster grows in size at the expense of small clusters.

Furthermore, $Ni_{44}Co_6Mn_{40}Sn_{10}$ also shows very similar thermo-magnetic behavior. As *M* versus *H/T* plots do not merge for temperatures between the spin freezing temperature and the MT temperatures, possibility of SPM nature of this alloy is ruled out. The frequency dependent AC susceptibility data show significant shift of maxima. The extracted parameters from the fits of frequency dependent peak of AC susceptibility data show that $Ni_{44}Co_6Mn_{40}Sn_{10}$ is indeed SSG below the blocking temperature. This result contradicts the SPM model that was suggested by Bhatti *et al* [29]. In contrast, $Ni_{44}Co_6Mn_{40}Sn_{10}$ shows weak AF ordering below 40 K, above which a weak FM ordering appears while $Ni_{45}Co_5Mn_{38}Sn_{12}$ alloy is AF below 180 K. This significant difference in the ground state of the matrix for these two alloys may possibly be attributed to their different *e/a* values [29]. Cross over from FM to AF ground state has been observed with increase in e/a ratio [34]. It is also established fact that MT temperature increases with increase in *e/a* ratio [66-67] for FSMA system. $Ni_{45}Co_5Mn_{38}Sn_{12}$ has higher *e/a* ratio value than that of $Ni_{44}Co_6Mn_{40}Sn_{10}$ which is also reflected in their comparative MT temperatures.

## 5. CONCLUSIONS

The current work delves into the contentious subject of magnetic ordering of the low temperature martensite phase in Co-doped Ni-Mn-Sn alloys and presents an in-depth account by combining complementary magnetotometry and SANS techniques. In particular, two specific compositions are chosen for this study: $Ni_{45}Co_5Mn_{38}Sn_{12}$ and $Ni_{44}Co_6Mn_{40}Sn_{10}$. Our results provide new insight that helps understand this complex issue. Major findings derived from this study are listed below:

1. Detailed magnetization and susceptibility studies confirm that the low temperature magnetic order in the martensite phase in $Ni_{45}Co_5Mn_{38}Sn_{12}$ is neither of SG nor of SPM in nature, rather of a long range AF type.

2. In contrast, $Ni_{44}Co_6Mn_{40}Sn_{10}$ exists in SSG state at low temperature, not in SPM state as reported earlier in the literature.

3. SANS study has revealed presence of nanometric clusters in $Ni_{45}Co_5Mn_{38}Sn_{12}$ that grow in size with increasing temperature and persist right up to martensitic transformation, ruling them out to be spin clusters.

4. These nanometric clusters are likely to be of structural origin and related to martensitic transformation.

5. Two different sizes of nanometric clusters (3-4 nm and 11-13 nm) are recorded, where the bigger clusters are seen to grow at the expense of the smaller ones.

## REFERENCES

1. Karaca H E, Karaman I, Basaran B, Ren Y, Chumlyakov Y I and Maier H J 2009 *Adv. Funct. Mater.* **19** 983.
2. Manosa L, Gonzalez-Alonso D, Planes A, Bonnot E, Barrio M, Tamarit J L, Aksoy S and Acet M 2010 *Nat. Mater.* **9** 478.
3. Krenke T, Duman E, Acet M, Wassermann E F, Moya X, Manosa L and Planes A 2005 *Nat. Mater.* **4** 450.
4. Kainuma R, Imano Y, Ito W, Sutou Y, Morito H, Okamoto S, Kitakami O, Oikawa K, Fujita A, Kanomata T and Ishida K 2006 *Nature* **439** 957.

5. Srivastava V, Song Y, Bhatti K P and James R D 2011 Adv. Energy Mater. **1** 97.
6. Han Z D, Wang D H, Zhang C L, Tang S L, Gu B X and Du Y W 2006 *Appl. Phys. Lett.* **89** 182507.
7. Koyama K, Okada H, Watanabe K, Kanomata T, Kainuma R, Ito W, Oikawa K and Ishida K 2006 *Appl. Phys. Lett.* **89** 182510.
8. Yu S Y, Liu Z H, Liu G D, Chen J L, Cao Z X, Wu G H, Zhang B and Zhang X X, 2006 *Appl. Phys. Lett.* **89** 162503.
9. Kainuma R, Imano Y, Ito W, Morito H, Sutou Y, Oikawa K, Fujita A, Ishida K, Okamoto S, Kitakami O and Kanomata T 2006 *Appl. Phys. Lett.* **88** 192513.
10. Yu S Y, Ma L, Li G D, Liu Z H, Chen J L, Cao Z X, Wu G H, Zhang B and Zhang X X 2007 *Appl. Phys. Lett.* **90** 242501.
11. Umetsu R Y, Ito K, Ito W, Koyama K, Kanomata T, Ishida K and Kainuma R 2011 *J. Alloys Compd.* **5** 1389.
12. Siruguri V, Babu P D, Kaushik S D, Biswas A, Sarkar S K, Krishnan M and Chaddah P 2012 *J. Phys.: Condens. Matter* **49** 496011.
13. Cong D Y, Roth S and Schultz L 2012 *Acta Mater.* **60** 5335.
14. Wang Y D, Huang E W, Ren Y, Nie Z H, Wang G, Liu Y D, Deng J N, Choo H, Liaw P K, Brown D E and Zuo L 2008 *Acta Mater.* **56** 913.
15. Srivastava V, Chen X and James R D 2010 *Appl. Phys. Lett.* **97** 014101.
16. Aksoy S, Acet M, Deen P P, Manosa L and Planes A 2009 *Phys. Rev. B* **79** 212401.
17. Planes A, Manosa L and Acet M 2009 *J. Phys.: Condens. Matter* **21** 233201.
18. Ma L, Wang W H, Lu J B, Li J Q, Zhen C M, Hou D L and Wu G H 2011 *Appl. Phys. Lett.* **99** 182507.
19. Chatterjee S, Giri S, De S K and Majumdar S 2009 *Phys. Rev. B* **79** 092410.
20. Singh N, Borgohain B, Srivastava A K, Dhar A and Singh H K 2016 *Appl. Phys. A* **122** 237.
21. Tian F, Cao K, Zhang Y, Zeng Y, Zhang R, Chang T, Zhou C, Xu M, Song X and Yang S 2016 Sci. Rep. **6** 30801.
22. Wang B M, Liu Y, Ren P, Xia B, Ruan K B, Yi J B, Ding J, Li X G and Wang L 2011 *Phys. Rev. Lett.* **106** 077203.
23. Umetsu R Y, Fujita A, Ito W, Kanomata T and Kainuma R 2011 *J. Phys.: Condens. Matter* **23** 326001.

24. Srivastava S K, Srivastava V K, Varga L K, Khovaylo V V, Kainuma R, Nagasako M and Chatterjee R 2011 *J. Appl. Phys.* **109** 083915.
25. Liao P, Jing C, Wang X L, Yang Y J, Zheng D, Li Z, Kang B J, Deng D M, Cao S X, Zhang J C and Lu B 2014 *Appl. Phys. Lett.* **104** 92410.
26. Agarwal S, Banerjee S and Mukhopadhyay P K 2013 *J. Appl. Phys.* **114** 133904.
27. Cong D Y, Roth S, Liu J, Luo Q, Potschke M, Hurrich C and Schultz L 2010 *Appl. Phys. Lett.* **96** 112504.
28. Perez-Landazabal J I, Recarte V, Sanchez-Alarcos V, Gomez-Polo C and Cesari E 2013 *Appl. Phys. Lett.* **102** 101908.
29. Bhatti K P, El-Khatib S, Srivastava V, James R D and Leighton C 2012 *Phys. Rev. B* **85** 134450.
30. Kumar D 2014 *J. Phys.: Condens. Matter* **26** 276001.
31. Binder K and Young A P 1986 *Rev. Mod. Phys.* **58** 801.
32. Mydosh J A 1993 *Spin Glasses: An Experimental Introduction (London: Taylor and Francis).*
33. Jing C, Yang Y, Yu D, Li Z, Wang X, Kang B, Cao S, Zhang J, Zhu J and Lu B 2014 *Adv. Mat. Res.* **875** 272.
34. Krenke T, Acet M, Wassermann E F, Moya X, Manosa L and Planes A 2005 *Phys. Rev. B* **72** 014412.
35. Krenke T, Acet M, Wassermann E F, Moya X, Manosa L and Planes A 2006 *Phys. Rev B* **73** 174413.
36. Wang B M, Liu Y, Ren P, Xia B, Ruan K B, Yi J B, Ding J, Li X G and Wang L 2011 *Phys. Rev. Lett.* **106** 077203.
37. Cong D Y, Roth S, Potschke M, Hurrich C and Schultz L 2010 *Appl. Phys. Lett.* **97** 021908.
38. Nogues J and Schuller I K 1999 *J. Magn. Magn. Mater.* **192** 203–232.
39. Nowak U, Usadel K D, Keller J, Miltenyi P and Guntherodt G 2002 *Phys. Rev. B* **66** 014430.
40. Venkateswarlu B, Krishnan R H, Chelvane J A, Babu P D and Kumar N H 2019 *J. Alloys Compd.* **777** 373-381.
41. Yasin S M, Saha R, Srinivas V, Kasiviswanathan S and Nigam A K 2016 *J. Magn. Magn. Mater.* **418** 158–162.
42. Banerjee S K 1964 *Phys. Lett.* **12** 16‑17.

43. Sahoo R, Nayak A K, Suresh K G and Nigam A K 2011 *J. Appl. Phys.* **109** 12390.
44. Bhatti I N, Rawat R, Banerjee A and Pramanik A K 2014 *J. Phys.: Condens. Mater*. **27** 016005.
45. Li M M, Shen J L, Wang X, Ma L, Li G K, Zhen C M, Hou D L and Wang M 2018 *Intermetallics* **96** 13–17.
46. Goya G F, Berquo T S, Fonseca F C and Morales M P 2003 *J. Appl. Phys.* **94** 3520.
47. Ma L, Wang W H, Lu J B, Li J Q, Zhen C M, Hou D L and Wu G H 2011 *Appl. Phys. Lett.* **99** 182507.
48. Thakur M, Patra M, Majumdar S and Giri S 2009 *J. Appl. Phys.* **105** 073905.
49. Malinowski A, Bezusyy V L, Minikayev R, Dziawa P, Syryanyy Y and Sawicki M 2011 *Phys. Rev. B* **84** 024409.
50. Mulder C A M, Duyneveldt A J V and Mydosh J A 1981 *Phys. Rev. B* **23** 1384‐1396.
51. Dormann J L, Bessais L and Fiorani D 1988 *J. Phys. C* **21** 2015.
52. Bedanta S and Kleemann W 2009 *J. Phys. D: Appl. Phys*. **42** 013001.
53. Parker D, Ladieu F, Vincent E, Meriguet G, Dubois E, Dupuis V and Perzynski R 2005 *J. Appl. Phys.* **97** 10A502.
54. Neel L 1949 *Ann. Geophys*. **5** 99.
55. Brown Jr. W F 1963 Phys. Rev. **130** 1677.
56. Blundell S 2001 *Magnetism in Condensed Matter, Oxford University Press*.
57. Pedersen J 1994 *J. Appl. Cryst.* **27** 595-608.
58. Guinier G F A, Walker B C and Yudowith L K 1955 *Small Angle Scattering of X-rays, Wiley, New York.*
59. Aitchison J 1957 *The Lognormal Distribution, Cambridge University Press.*
60. Weissmueller J, Michels A, Barker J G, Wiedenmann A, Erb U and Shull R D 2001 *Phys. Rev. B* **63** 214414.
61. Deschamps A, Militzer M and Poole W J 2001 *ISIJ International* **41** 196–205.
62. Borgohain B, Siwach P K, Singh N and Singh H K 2018 *J. Magn. and Magn. Mat.* **454** 13–22.
63. Chernenko V A, Kakazei G N, Perekos A O, Cesari E and Besseghini S 2008 *J. Magn. Magn. Mater.* **320** 1063.
64. Liu D M, Nie Z H, Wang G, Wang Y D, Brown D E, Pearson J, Liaw P K and Ren Y 2010 *Mater. Sci. Eng. A* **527** 3561.
65. Chatterjee S, Giri S, Majumdar S and De S K 2008 *Phys. Rev. B* **77** 224440.

66. Sarkar S K, Biswas A, Babu P D, Kaushik S D, Srivastava A, Siruguri V and Krishnan M 2013 *J. Alloys Comp.* **586** 515-523.
67. Sarkar S K, Sarita, Babu P D, Biswas A, Siruguri V and Krishnan M 2016 *J. Alloys Comp.* **670** 281-288.

## List of Tables

| Sample | Ni | Co | Mn | Sn | e/a |
|---|---|---|---|---|---|
| $Ni_{45}Co_5Mn_{38}Sn_{12}$ | 44.3 ± 0.24 | 5.1 ± 0.11 | 38.1 ± 0.28 | 12.5 ± 0.2 | 8.06 |
| $Ni_{44}Co_6Mn_{40}Sn_{10}$ | 43.7± 0.1 | 6.2 ± 0.3 | 39.9 ± 0.2 | 10.2 ± 0.3 | 8.12 |

**Table 1.** EDX results showing chemical composition of $Ni_{45}Co_5Mn_{38}Sn_{12}$ and $Ni_{44}Co_6Mn_{40}Sn_{10}$ alloys with standard deviation while last column shows e/a value calculated for the alloys.

| Sample | $M_s$ (K) | $M_f$(K) | $A_s$(K) | $A_f$(K) | $T_{CM}$(K) | $T_{CA}$(K) |
|---|---|---|---|---|---|---|
| $Ni_{45}Co_5Mn_{38}Sn_{12}$ | 297 K | 264 | 280 | 308 | 225 | 375 |
| $Ni_{44}Co_6Mn_{40}Sn_{10}$ | 421 | 389.5 | 402.1 | 435 | 175 | 445 |

**Table 2.** Martensitic transition temperatures of the alloys as obtained from DSC scans.

| Model | Parameters | $Ni_{45}Co_5Mn_{38}Sn_{12}$ | $Ni_{44}Co_6Mn_{40}Sn_{10}$ |
|---|---|---|---|
| Mydosh | $\Phi$ | 0.0017 | 0.027 |
| Neel-Arrhenius | $\tau_0$(s) | $1.58\times10^{-99}$ | $3.6\times10^{-46}$ |
| | $E_a/K_B$(K) | $2.29\times10^{5}$ | $1.08\times10^{4}$ |
| Vogel-Fulcher | $\tau_0$(s) | $8.24\times10^{-46}$ | $1.93\times10^{-8}$ |
| | $E_a/K_B$(K) | $1.30\times10^{3}$ | 150.7 |
| | $T_P$(K) | 168.3 | 104.1 |
| Critical slowing down | $\tau_0$(s) | $8.34\times10^{-43}$ | $6.3\times10^{-13}$ |
| | $z\upsilon$ | 249.7 | 10.7 |
| | $T_0$(K) | 168.5 | 104 |

**Table 3.** Physical dynamic scaling parameters as obtained from the fitting of the experimental frequency dependent AC susceptibility peaks near $T_f$ using equations (1)-(4) for $Ni_{45}Co_5Mn_{38}Sn_{12}$ and $Ni_{44}Co_6Mn_{40}Sn_{10}$ alloys.

| Sample | Mydosh | CSD par. | | V-F par. | | | Comments |
|---|---|---|---|---|---|---|---|
| | $\Phi$ | $\tau_0$ (S) | $z\upsilon$ | $T_0$ (K) | $\tau_0$ (s)/$\omega_0$ (rad/s) | $E_a/K_B$ (K) | |
| $Ni_{1.6}Mn_2Sn_{0.4}$ bulk | 0.023 | | | 113 | $2\times10^{6}$rad/s | 75 | RSG [18] |
| $Ni_2Mn_{1.36}Sn_{0.64}$ | 0.06 | | | 81.5 | $3.14\times10^{6}$rad/s | 100.9 | RSG [19] |
| $Ni_{1.6}Mn_2Sn_{0.4}$ ribbon | 0.014 | $1.3\times10^{-13}$ | 6.95 | 122 | $1.3\times10^{-9}$s | 154.3 | SSG [20] |
| $Ni_{50}Mn_{38}Ga_{12}$ | | | | 65.1 | $9.3\times10^{-14}$s | 43.8 | SSG [21] |
| $Ni_{50}Mn_{38}Ga_{11}Sb_1$ | | | | 90.3 | $8.7\times10^{-12}$s | 72.5 | SSG [21] |
| $Ni_{50}Mn_{38}Ga_{10}Sb_2$ | | | | 130.7 | $2.4\times10^{-10}$s | 104.1 | CSG [21] |
| $Ni_{50}Mn_{38}Ga_9Sb_3$ | | | | 140.4 | $5.3\times10^{-9}$s | 158.4 | CSG [21] |
| $Ni_{50}Mn_{38}Ga_8Sb_4$ | | | | 147.5 | $7.2\times10^{-9}$s | 173.3 | CSG [21] |
| $Ni_{50}Mn_{38}Ga_7Sb_5$ | | | | 151.9 | $2.1\times10^{-8}$s | 191.7 | CSG [21] |
| $Ni_{50}Mn_{38}Ga_6Sb_6$ | | | | 156.4 | $1.3\times10^{-7}$s | 213.6 | CSG [21] |
| $Ni_{50}Mn_{37}In_{13}$ | | $10^{-8}$s | 9.7 | | | | SSG [22] |
| $Ni_{50}Mn_{40}Sb_{10}$ | 0.002 | $10^{-15}$s | 5.5 | | | | SSG [23] |
| $Ni_{50}Mn_{36}Co_4Sn_{10}$ | 0.034 | $10^{-9}$s | 5.9 | | | | SSG [25] |
| $Ni_{50}Mn_{35}In_{15}$ | 0.009 | $10^{-12}$s | 5.5 | | | | CSG [23] |
| $Ni_{50}Mn_{38.5}Sn_{11.5}$ | 0.007 | $10^{-13}$s | 5.5 | | | | CSG [23] |
| $Ni_{50}Mn_{34}Sn_6Al_{10}$ | 0.094 | $1.2\times10^{-5}$s | 5.5 | | $1.9\times10^{14}$rad/s | 3000 | CSG [26] |
| $Ni_{43.5}Co_{6.5}Mn_{39}Sn_{11}$ | 0.024 | $10^{-12}$s | 9.7 | | | | SSG [27] |
| $Ni_{50}Mn_{39}Sn_{11}$ | 0.0128 | $10^{-12}$s | 7.0 | | | | SSG below $T_f$ [13] |
| $Ni_{45}Co_5Mn_{39}Sn_{11}$ | 0.0171 | $10^{-12}$s | 8.7 | | | | SSG below $T_f$ [13] |
| $Ni_{38}Mn_{48}Cr_3Sn_{11}$ | 0.028 | $2\times10^{-19}$s | | 9 | $1.24\times10^{-15}$s | 966 | CSG [60] |
| $Ni_{38}Mn_{49}Cr_2Sn_{11}$ | 0.0093 | $10^{-43}$s | | 109 | $2.6\times10^{-11}$s | 498.7 | SSG [60] |

**Table 4.** Physical dynamic scaling parameters for various Ni-Mn-based FSMAs showing various magnetic ground state for different alloys as available in literature.

## Figures

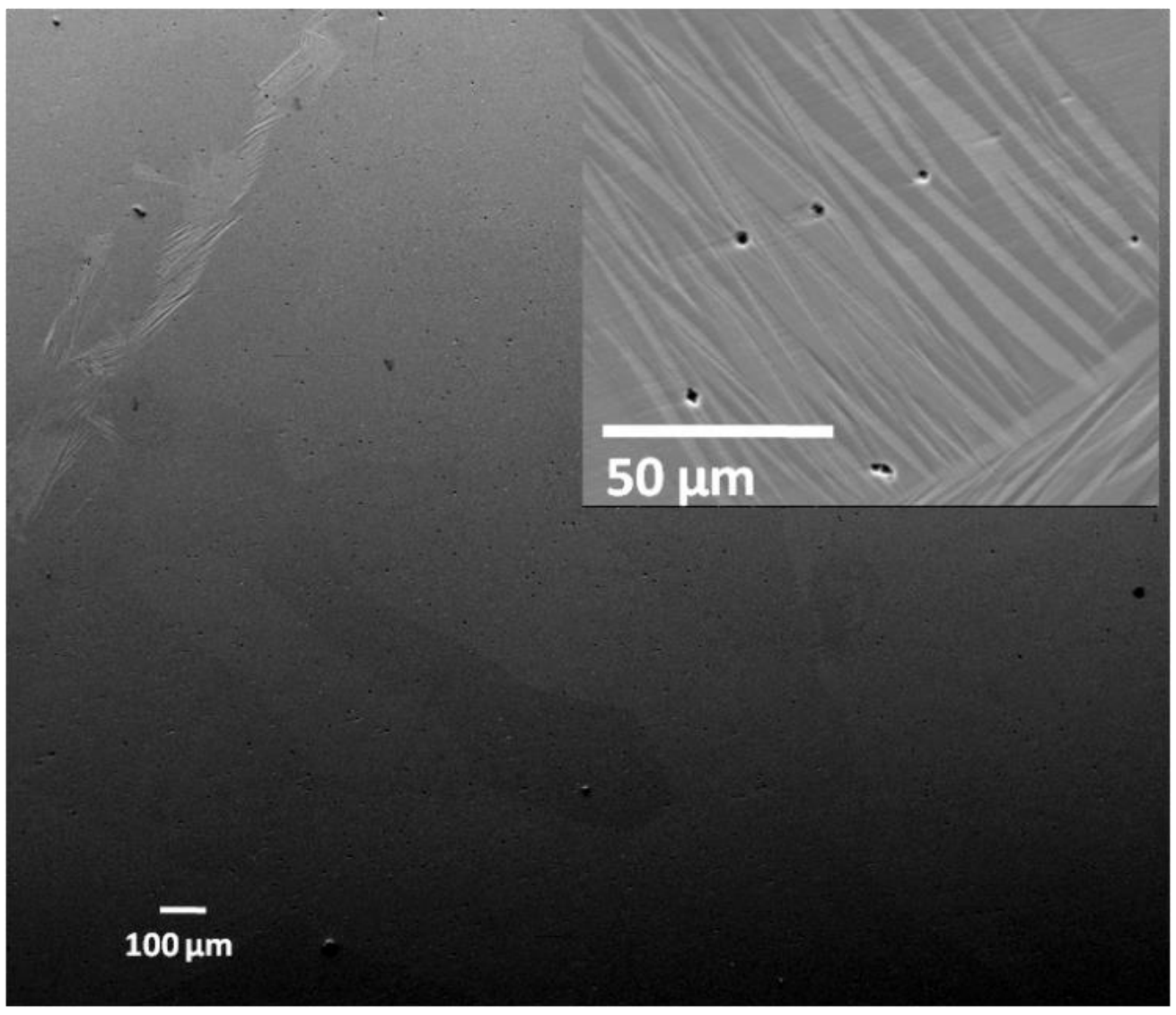

**Figure 1.** Representative SEM image for $Ni_{45}Co_5Mn_{38}Sn_{12}$ alloy showing predominantly austenite matrix with fraction of martensite. The inset shows zoomed portion of twinned martensite (upper right area in zoomed image).

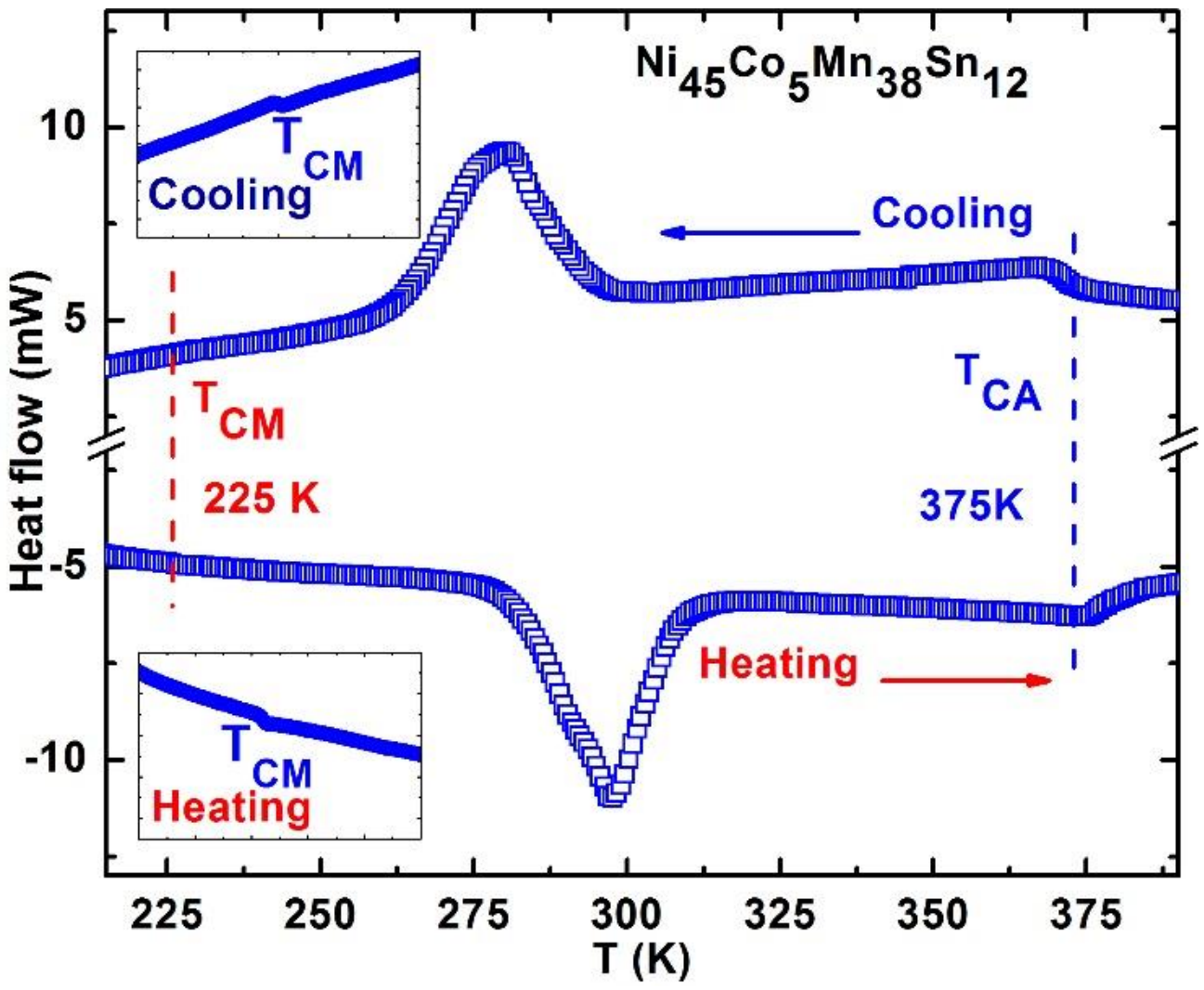

**Figure 2.** DSC plot showing characteristic martensitic transformation temperatures for $Ni_{45}Co_5Mn_{38}Sn_{12}$ alloy.

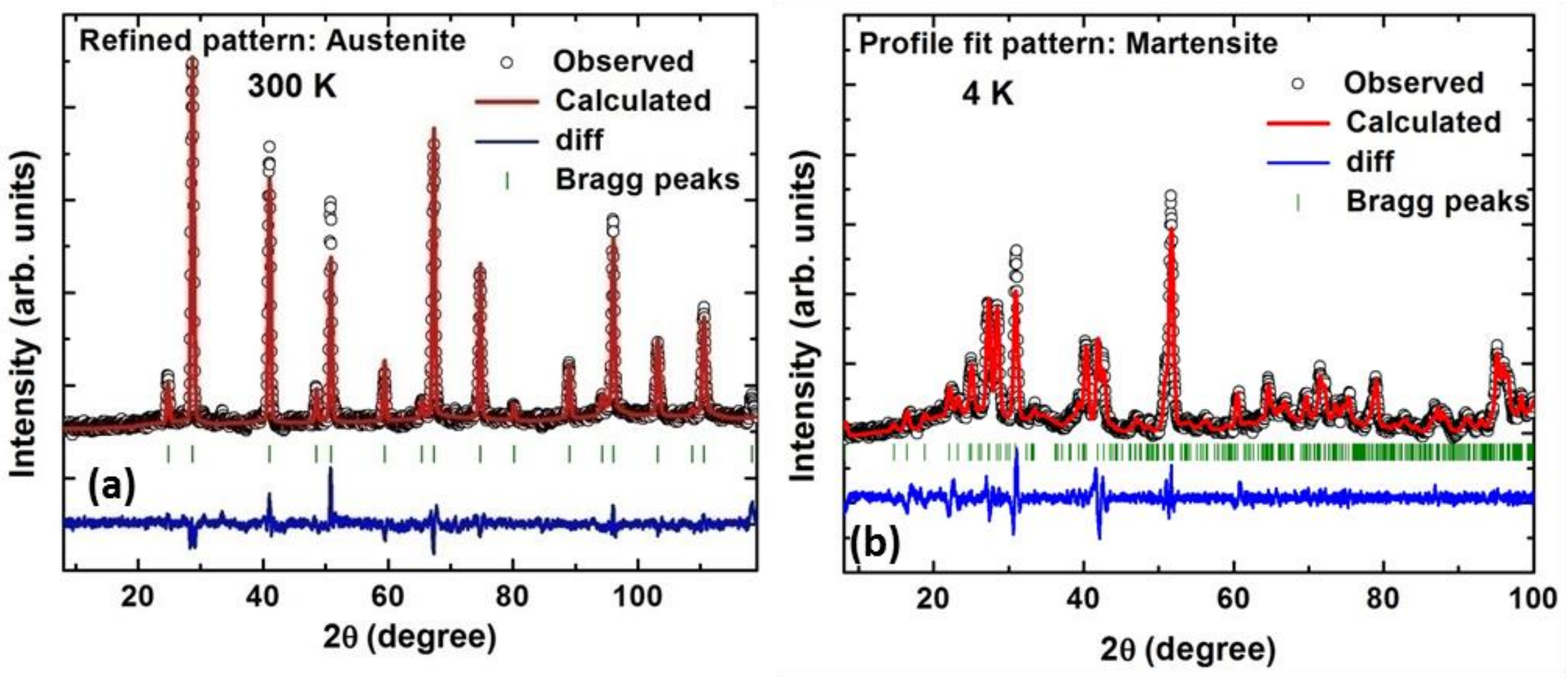

**Figure 3.** Le Bail fitted ND pattern for $Ni_{45}Co_5Mn_{38}Sn_{12}$ alloy at: (a) RT and (b) 4 K.

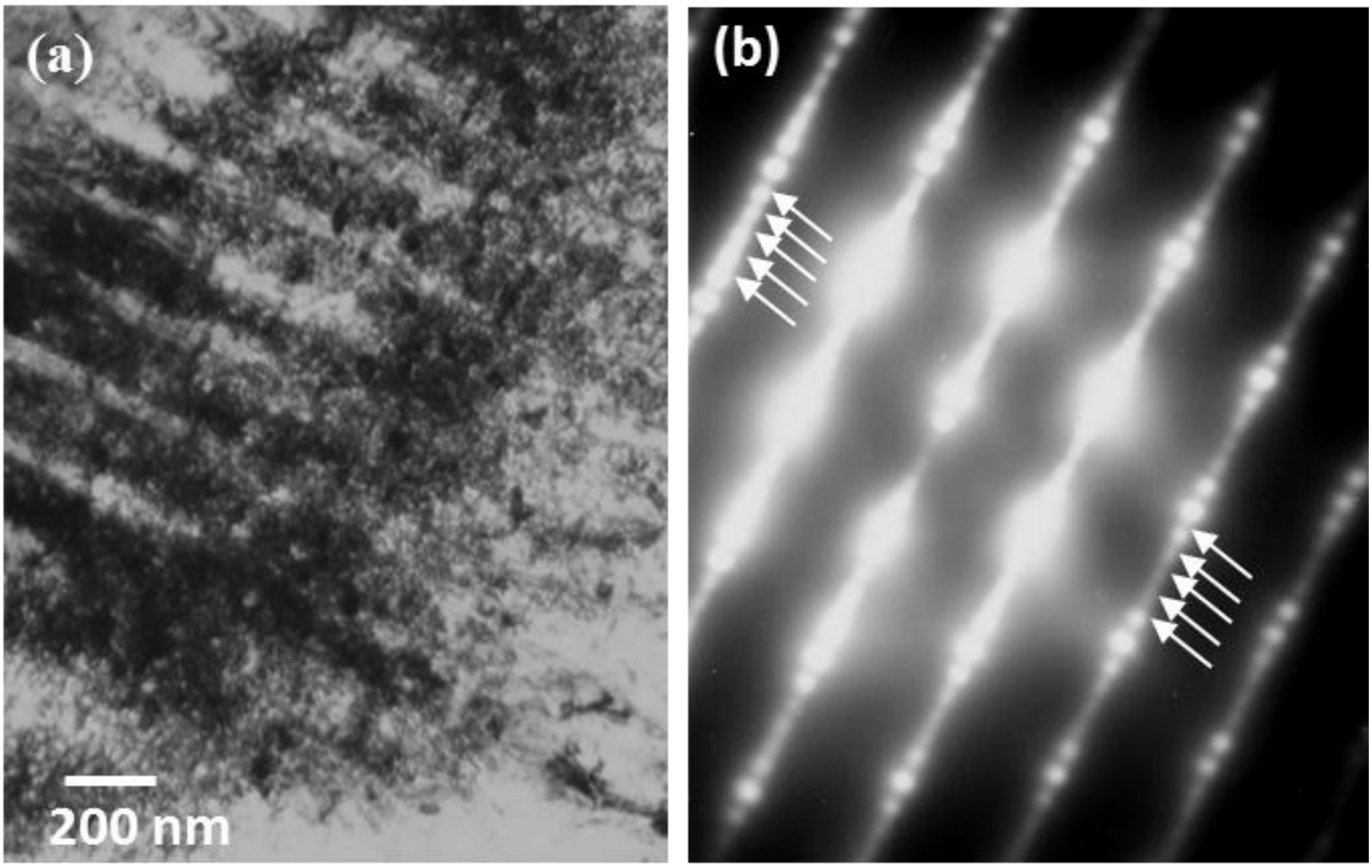

**Figure 4.** TEM micrographs of $Ni_{45}Co_5Mn_{38}Sn_{12}$ alloy: (a) bright field image shows martensite plates; (b) Corresponding SAED shows the satellite spots as evidence of 6M.

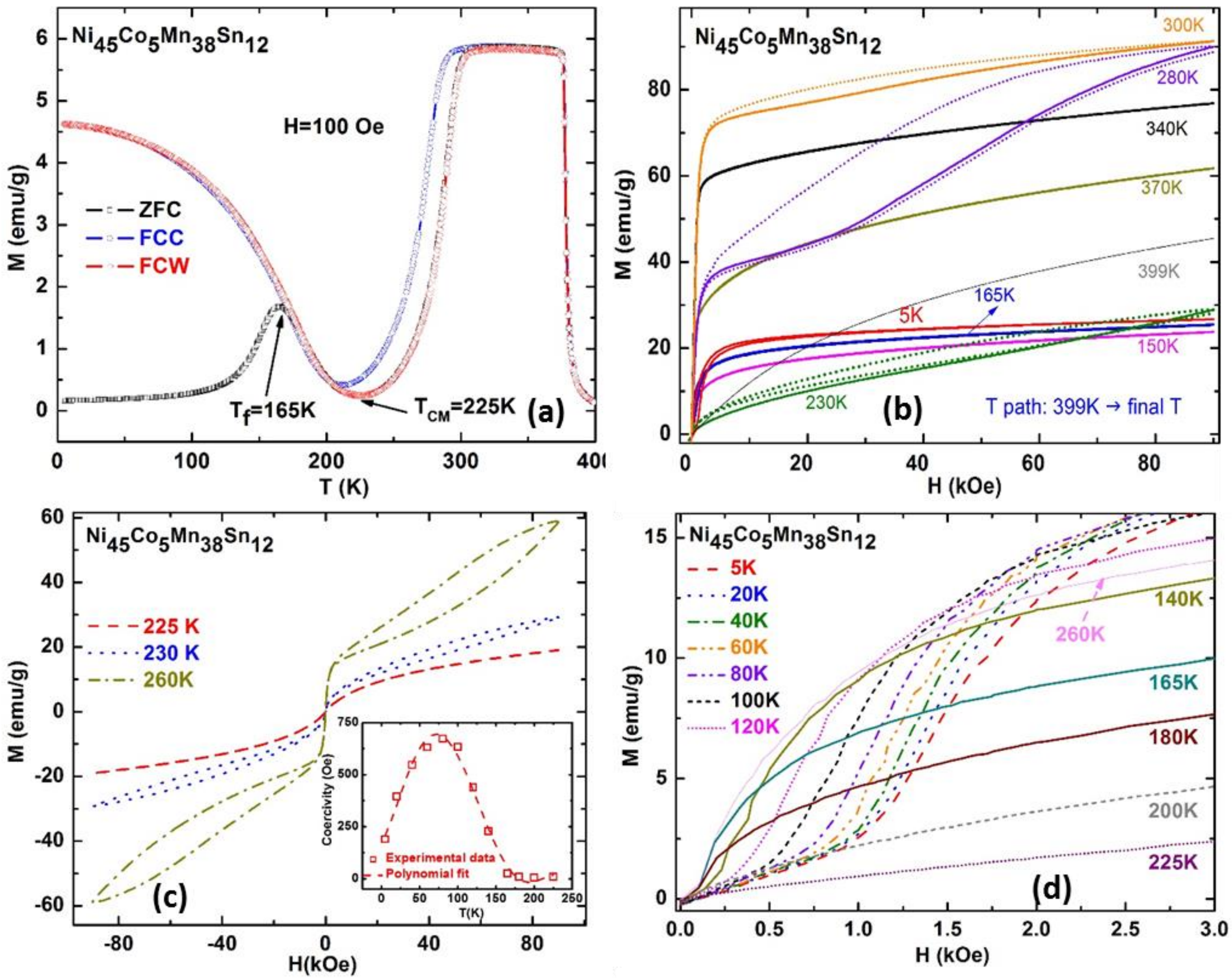

**Figure 5.** For $Ni_{45}Co_5Mn_{38}Sn_{12}$ alloy: (a) *M* versus *T* plot in ZFC, FCC and FCW for applied magnetic field of 100 Oe, (b) representative *M* versus *H* curves measured during cooling cycle from 399 K to 5 K, (c) non linearity with increasing field is evident from *M* versus *H* curves above $T_f$ temperature, while inset showing variation of coercivity with temperature, (d) virgin low field *M* versus *H* plot indicating magnetic ordering above and below $T_f$ temperature.

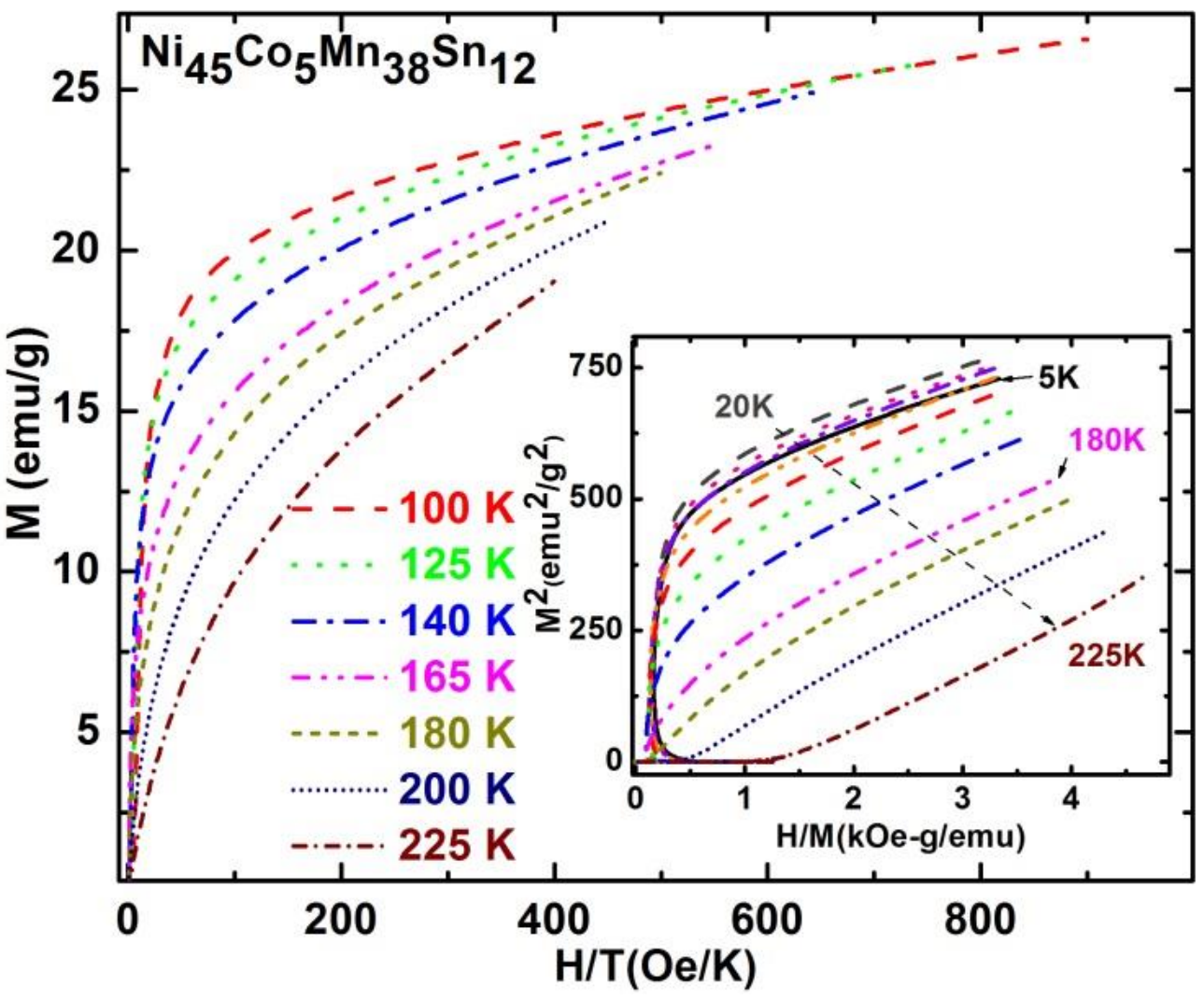

**Figure 6.** *M* versus *H/T* plot for $Ni_{45}Co_5Mn_{38}Sn_{12}$ alloy while inset showing $M^2$ versus *H/M* variation.

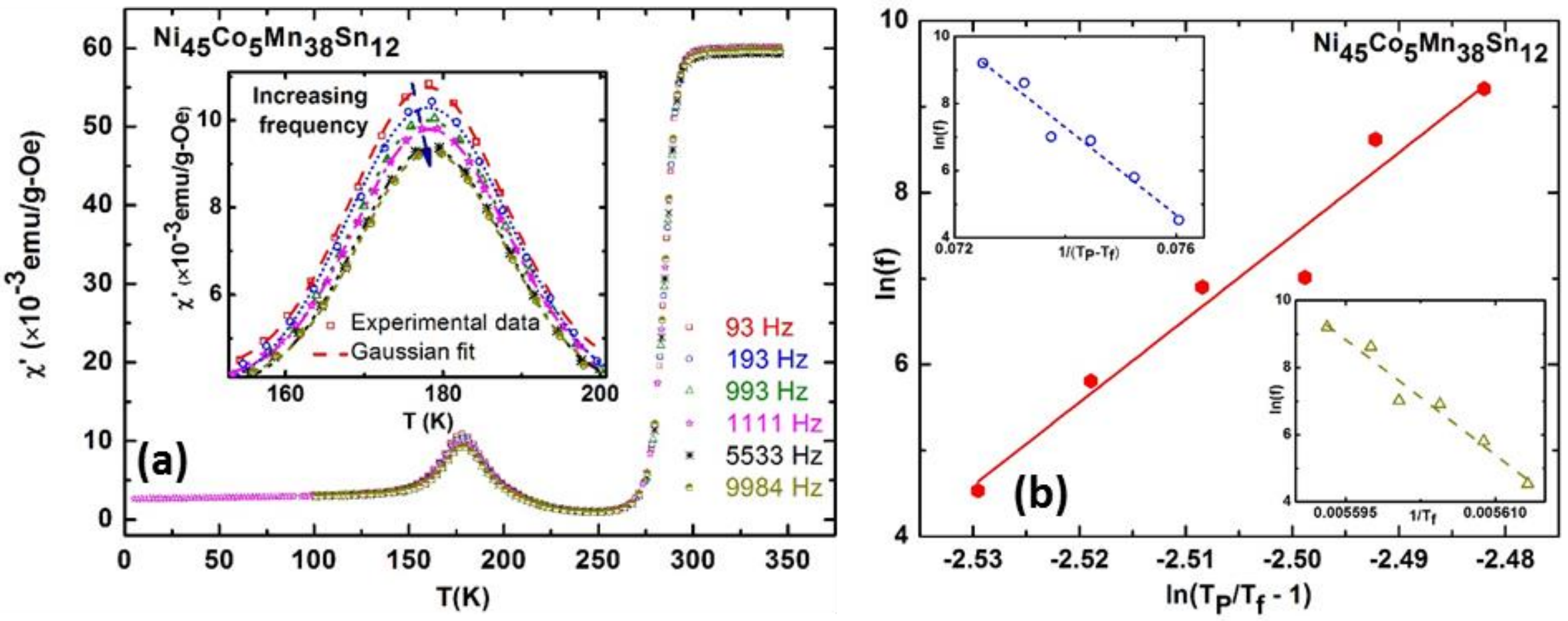

**Figure 7.** (a) Temperature dependence of real part of AC susceptibility for $Ni_{45}Co_5Mn_{38}Sn_{12}$ alloy measured at different frequencies from 93 to 9984 Hz while inset showing the frequency dependence of the peak as observed in $\chi'$ versus $T$ plot near $T_f$, (b) plot for ln($f$) versus ln($T_P/T_f$ -1) (solid hexagon) with best fit to equation (2) (solid line), while upper inset showing variation of ln($f$) versus 1/($T_P$-$T_f$) (open circle) with best fit to equation (3) (dashed line) and lower inset shows for ln($f$) versus 1/($T_f$) (open triangle) with best fit to equation (4) (dashed line).

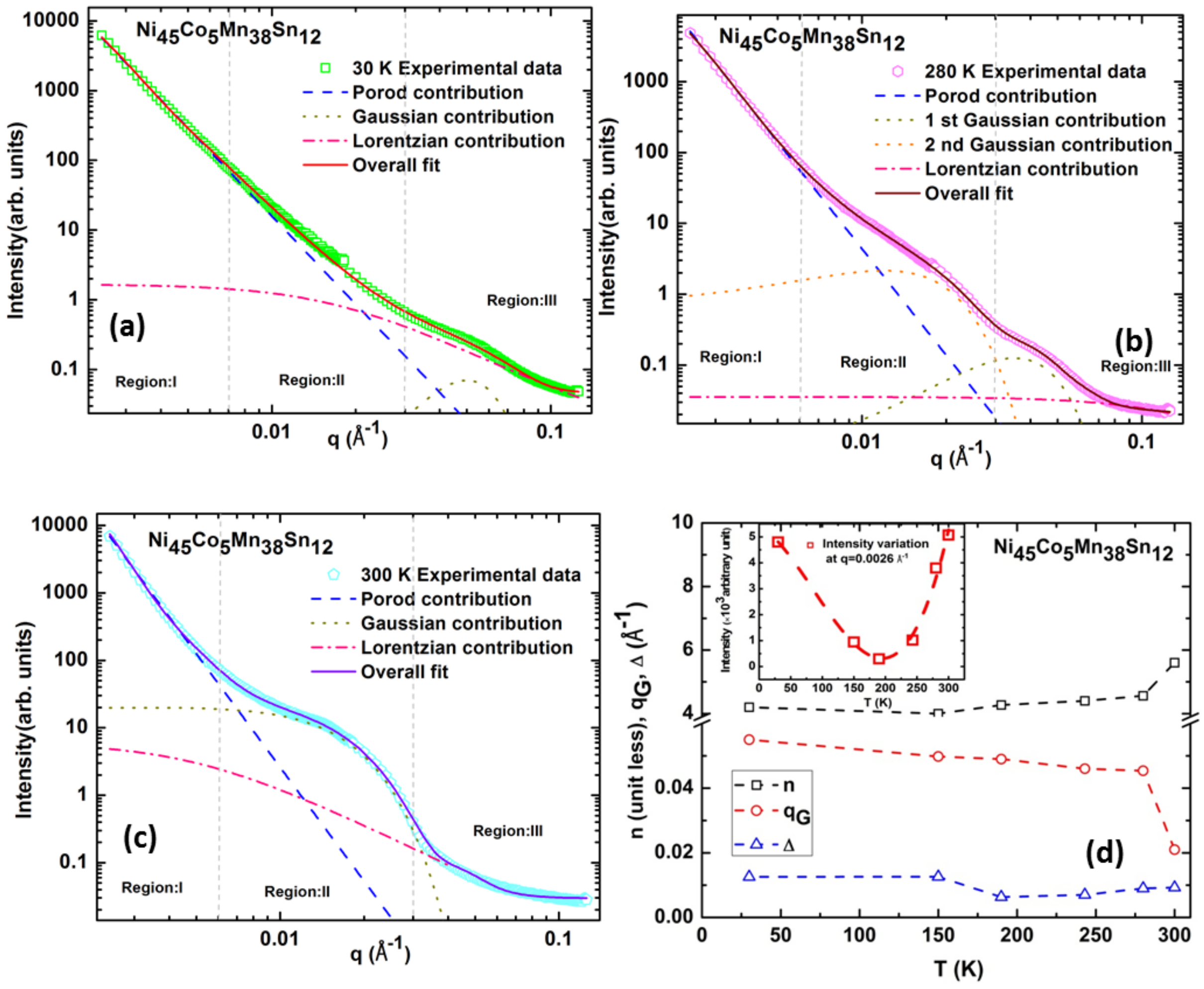

**Figure 8.** (a)-(c) Fitted SANS profile for $Ni_{45}Co_5Mn_{38}Sn_{12}$ alloy at three key temperatures 30 K, 280 K and 300 K respectively. The heavy solid line is the overall fit, while the dashed, dotted, and dashed dotted represent the individual Porod, Gaussian, and Lorentzian contributions. 280 K data requires two distinct Gaussian to fit the data. (d) Temperature variation of $n$, $q_G$ and $\Delta$, as extracted from fits while inset shows variation of scattering intensity with temperature at selected $q$ value of 0.0026 $Å^{-1}$, consistent with magnetic behaviour of the alloy.

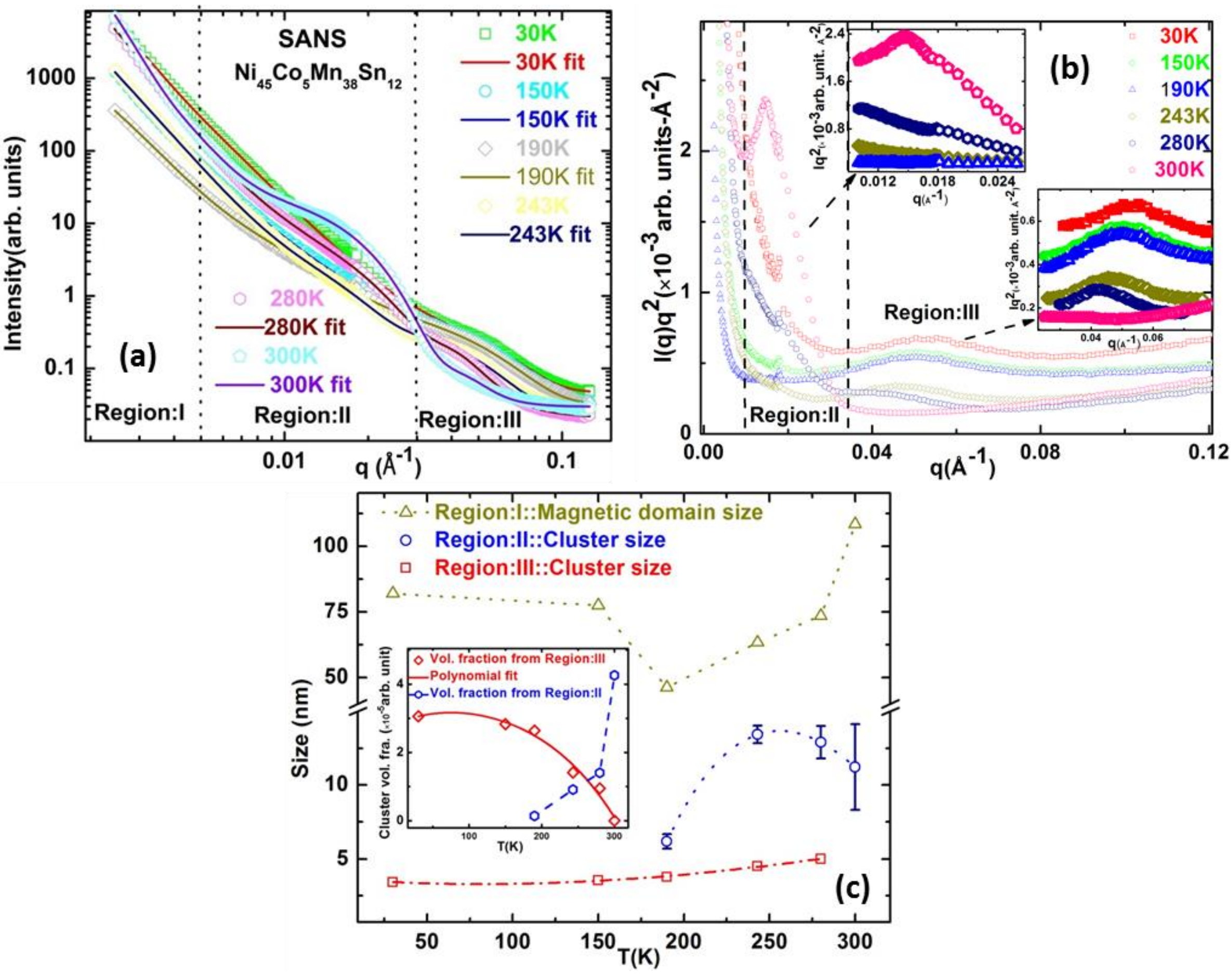

**Figure 9.** (a) SANS profile considering spherical clusters for $Ni_{45}Co_5Mn_{38}Sn_{12}$ alloy for several key temperatures (30-300 K) with fits (solid lines). (b) $I(q)q^2$ versus $q$ plot, upper inset showing increasing volume fraction from region-II while lower inset shows decreasing volume fraction from region-III for the involved nano-clusters. (c) Temperature variations of the extracted physical parameters from SANS analysis.

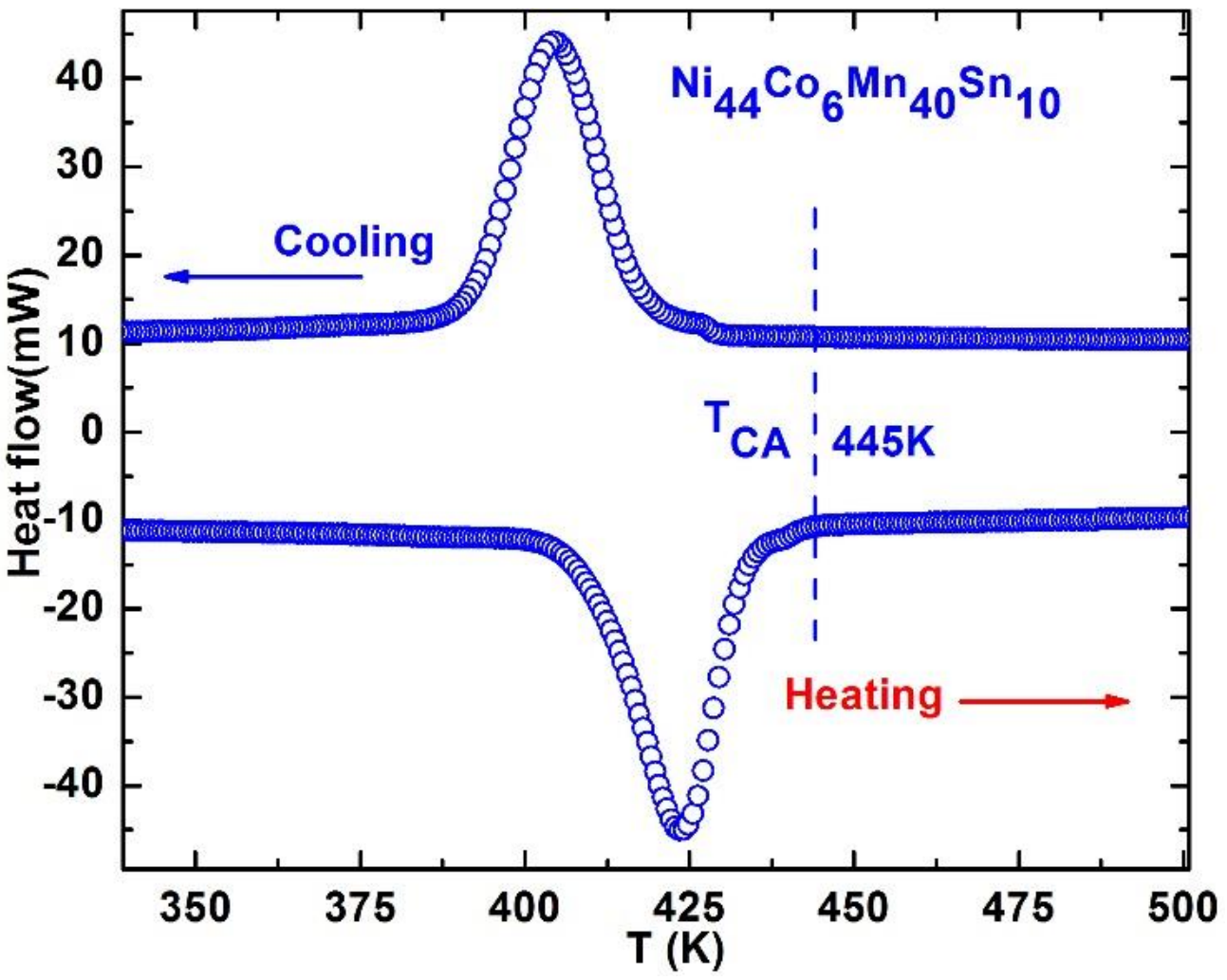

**Figure 10.** DSC plot showing characteristic martensitic transformation temperatures for $Ni_{44}Co_6Mn_{40}Sn_{10}$ alloy.

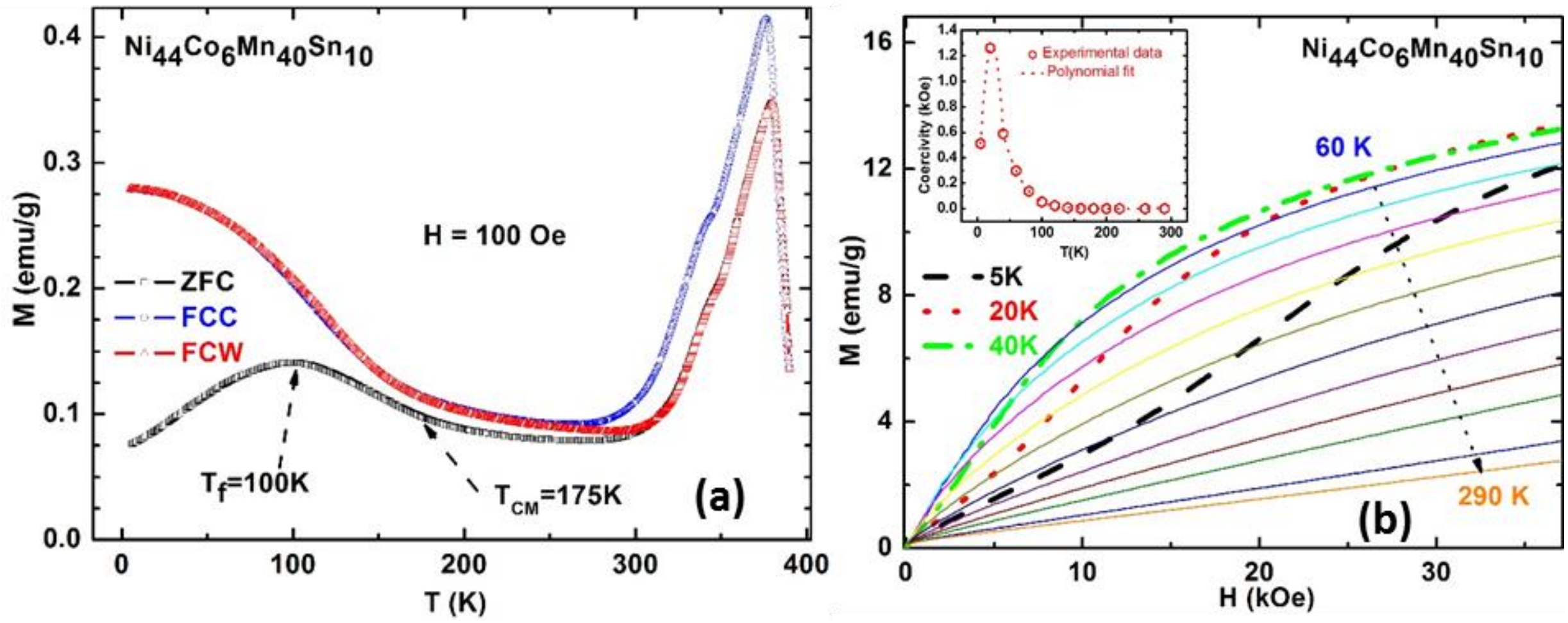

**Figure 11.** (a) *M* versus *T* plot for $Ni_{44}Co_6Mn_{40}Sn_{10}$ alloy in ZFC, FCC, FCW sequences for applied magnetic field of 100 Oe, (b) representative virgin low field *M* versus *H* plot from 290 K to 5 K, the highlighted isotherms at 5 K, 20 K and 40 K show AF ordering while at high temperature FM resides. The inset shows temperature variation of coercivity.

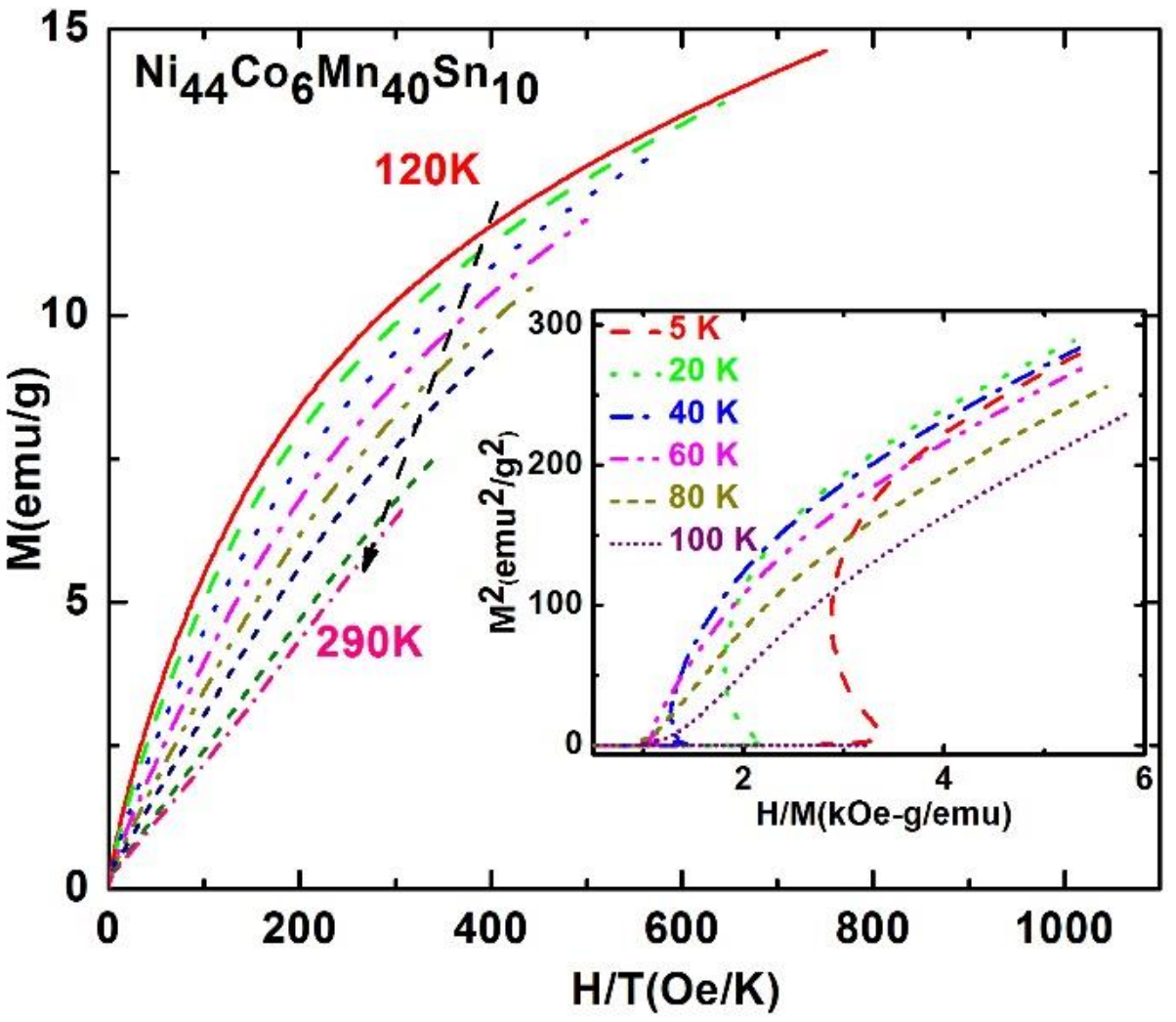

**Figure 12.** *M* versus *H/T* plot for $Ni_{44}Co_6Mn_{40}Sn_{10}$ alloy, while inset showing $M^2$ versus *H/M* plot.

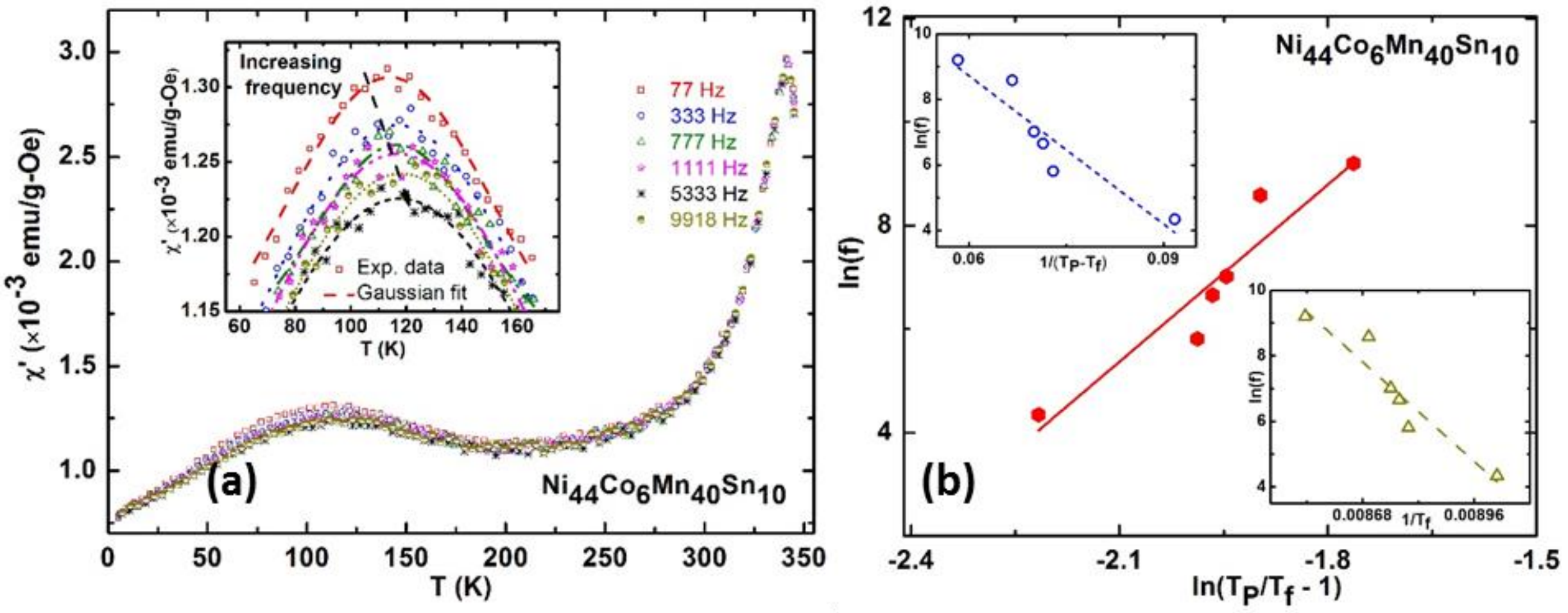

**Figure 13.** (a) Temperature dependence of real part of AC susceptibility for $Ni_{44}Co_6Mn_{40}Sn_{10}$ alloy measured at different frequencies from 77 to 9918 Hz while inset showing the frequency dependence of the peak as observed in $\chi'$ versus *T* plot near $T_f$, (b) plot for ln(*f*) versus ln($T_P/T_f$ -1) (solid hexagon) with best fit to equation (2) (solid line), while upper inset showing variation of ln(*f*) versus 1/($T_P$-$T_f$) (open circle) with best fit to equation (3) (dashed line) and lower inset shows for ln(*f*) versus 1/($T_f$) (open triangle) with best fit to equation (4) (dashed line).